\documentclass[journal]{IEEEtran}
\IEEEoverridecommandlockouts
\usepackage{times,amsmath,amsfonts,amssymb,amsthm}
\usepackage{cite}
\usepackage{graphicx,subfigure,psfrag}
\usepackage{algorithm}
\usepackage{booktabs}
\usepackage{multirow}
\usepackage{multicol}
\usepackage{colortbl}
\usepackage{color}
\usepackage{algpseudocode}
\usepackage{balance}

\def\abf{{\bf a}}
\def\bbf{{\bf b}}
\def\cbf{{\bf c}}
\def\dbf{{\bf d}}

\def\gbf{{\bf g}}

\def\pbf{{\bf p}}

\def\Ac{{\cal A}}

\def\Dc{{\cal D}}

\def\Ic{{\cal I}}

\def\Kc{{\cal K}}

\def\Nc{{\cal N}}

\def\Qc{{\cal Q}}

\def\Sc{{\cal S}}
\def\Tc{{\cal T}}
\def\Uc{{\cal U}}

\def\Wc{{\cal W}}

\def\eg{{\it e.g.,\ \/}}

\def\ie{{\it i.e.,\ \/}}
\def\nn{\nonumber}

\def\Unique{\text{Unique}}
\def\sSc{\text{\tiny $\Sc$}}

\theoremstyle{definition}
\newtheorem{lemma}{Lemma}
\newtheorem{theorem}{Theorem}
\newtheorem{Remark}{Remark}
\newtheorem{proposition}{Proposition}
\newtheorem{myDef}{Definition}
\newtheorem{corollary}{Corollary}

\newtheorem*{assumption}{C1)}

\begin{document}

\title{Memory-Rate Tradeoff for Caching with Uncoded Placement under Nonuniform Random Demands
}
\author{Yong Deng, \IEEEmembership{Member, IEEE}  and Min~Dong, \IEEEmembership{Senior Member, IEEE}
\thanks{This work was supported by the Natural Sciences and Engineering
Research Council of Canada (NSERC) under the Discovery Grant. Partial results in this work were presented in~\cite{Deng2020Memory} and~\cite{Yong2020Optimal}. (Corresponding author: Min Dong.) }
\thanks{Yong Deng was with the Department of Electrical, Computer and Software Engineering, Ontario Tech University, Oshawa, ON L1G 0C5, Canada. He is now with the Edward S. Rogers Sr. Department of Electrical and Computer Engineering,
University of Toronto, Toronto, ON M5S 3G4, Canada (email:
yong.deng@utoronto.ca).}
\thanks{Min Dong is with the Department of Electrical, Computer and Software Engineering, Ontario Tech University, Oshawa, ON L1G 0C5, Canada (e-mail: min.dong@ontariotechu.ca).}}

\maketitle

\begin{abstract}
This paper considers a caching system of a single server and multiple users. We aim to characterize the memory-rate tradeoff for caching with  uncoded cache placement, under nonuniform file popularity. Focusing on the modified coded caching scheme (MCCS) recently proposed by Yu, \emph{etal.}, we formulate the cache placement  optimization problem for the MCCS to minimize the average delivery rate under nonuniform file popularity, restricting  to a class of popularity-first placements.
We then present two information-theoretic lower bounds on the average rate for caching with  uncoded placement, one  for general cache placements and the other restricted to the popularity-first  placements.
By comparing the average rate of the optimized MCCS with the lower bounds, we prove that the optimized MCCS attains the general lower bound for the two-user case, providing the exact memory-rate tradeoff. Furthermore, it attains the popularity-first-based lower bound for the case of general $K$ users with distinct file requests. In these two cases, our results also reveal that the popularity-first placement is optimal for the MCCS, and  zero-padding used in coded delivery incurs no loss of optimality. For the case of $K$ users with redundant file requests, our analysis shows that there may exist a gap between the optimized MCCS and the lower bounds due to zero-padding.
We next fully characterize the optimal popularity-first cache placement for the MCCS, which is shown to possess a simple file-grouping structure and can be computed via an efficient algorithm using closed-form expressions.
Finally, we extend our study to
accommodate nonuniformity in both file popularity and size, where we show that the optimized MCCS  attains the lower bound for the  two-user case, providing the exact memory-rate tradeoff. Numerical results show that, for general settings, the gap  between the optimized MCCS and the lower bound only exists in limited cases and is very small.     
\end{abstract}


\section{Introduction}
Caching has emerged as a key  technology  in future wireless networks for fast content distribution. By  distributively storing partial contents  near end-users at off-peak times,~it alleviates network traffic  and ensures timely delivery~\cite{Bastug&etal:COMMag14,Wang&etal:COMMag14,paschos2018role}.
For a system with a central server connecting to multiple cache-equipped users, the seminal work in \cite{Maddah-Ali&Niesen:TIT2014}  introduced a coded caching scheme (CCS) that combines carefully designed  uncoded cache contents
with a coded multicast delivery strategy. It was shown to provide a global caching gain in addition to local caching gain to reduce the delivery rate (load) substantially, leading to the possibility of serving an infinite number of users simultaneously with finite resources. This advantage has drawn intense interests in coded caching, with  extension to various transmission scenarios and network architectures, including decentralized scenarios \cite{Niesen&Maddah-Ali:TIT2015},
transmitter caching in mobile edge networks\cite{shariatpanahi2016multi,Sengupta&etal:TIT17,Maddah-Ali&etal:ISIT},
user caching in device-to-device networks\cite{ji2016fundamental},
 transmitter and receiver caching in wireless interference networks \cite{Naderializadeh&Maddah-Ali:TIT2017,Xu&etal:TIT17},
hierarchical networks\cite{karamchandani2016hierarchical},
and online caching \cite{pedarsani2016online}.
Recently, a modified coded caching scheme (MCCS) was proposed \cite{Yu&Maddah-Ali:TIT2018} with an improved delivery strategy to remove redundancy among coded messages, resulting in further  delivery rate reduction from that of the CCS.
The MCCS was  further   applied to the device-to-device  networks~\cite{yapar2019optimality}.

 For understanding the fundamental limit of   coded caching, many research efforts were devoted to characterizing the memory-rate tradeoff  for caching with uncoded cache placement.
For files of uniform popularity and sizes, this tradeoff has been  studied extensively, typically by developing an achievable scheme and compare it to an information-theoretic lower bound~\cite{Maddah-Ali&Niesen:TIT2014,Yu&Maddah-Ali:TIT2018,Wan2016On1,Wan2016On2}.
When the system has fewer users than files, it has been shown that the CCS with optimized cache placement achieves the minimum peak delivery rate   for  caching with uncoded placement, \ie the exact memory-rate tradeoff~\cite{Wan2016On1,Wan2016On2}. For  general scenarios  of  arbitrary users and files with random requests, the MCCS with optimized cache placement has been shown to characterize the exact memory-rate tradeoff that minimizes both average and peak delivery rate~\cite{Yu&Maddah-Ali:TIT2018} under uniform file popularity and sizes.

For  files with nonuniform popularity or sizes,  different cache placement strategies were proposed for the CCS to handle nonuniformity~\cite{Niesen&Maddah-Ali:TIT2017,hachem2017coded,Ji&Order:TIT17,Zhang&Coded:TIT18,Jin&Cui:Arxiv2017,Saberali&Lampe:TIT20,Deng&Dong:TIT22,Cheng&Li17Optimal,Zhang&Lin15Coded,Zhang&Lin19Closing,Daniel&Yu:TIT19}.
In particular, several   CCS-based schemes, either for nonuniform file popularity~\cite{Ji&Order:TIT17,Zhang&Coded:TIT18} or for~nonuniform file sizes\cite{Zhang&Lin15Coded,Zhang&Lin19Closing}, were shown to achieve an average rate that is a constant factor away from the lower bound for caching with any cache placement. However,  these gaps are still large for practical concerns. Only recently, for two files of nonuniform popularity, a coded caching scheme was proposed~\cite{Sahraei2019TheOptimal}, which achieves the lower bound for caching with uncoded placement. For the MCCS,  existing studies are scarce, and only \cite{Jin&Cui:Arxiv2018} studied the cache placement  optimization under nonuniform file popularity. In general, for files with nonuniform popularity and sizes, characterizing the memory-rate tradeoff  for caching with uncoded placement is challenging. How well the CCS and the MCCS perform in terms of  memory-rate tradeoff  under  nonuniform file popularity or size remains unknown.

The cache placement is a key design issue for coded caching to maximize  caching gain
and minimize  delivery rate. For uniform file popularity, a symmetric cache placement (\ie identical cache placement for all files) is optimal for both the CCS~\cite{Daniel&Yu:TIT19} and the MCCS~\cite{Yu&Maddah-Ali:TIT2018}.
For nonuniform file popularity, the problem is much more challenging, as
the cache placement may
be asymmetric among files. This asymmetry introduces nonequal subfile sizes that complicate both  design and analysis.
For the CCS,  suboptimal placement strategies ~\cite{Niesen&Maddah-Ali:TIT2017,Ji&Order:TIT17,hachem2017coded,Zhang&Coded:TIT18}  and optimization approaches~\cite{Daniel&Yu:TIT19,Jin&Cui:Arxiv2017,Saberali&Lampe:TIT20,Deng&Dong:TIT22} were proposed to study the cache placement. In particular, the  optimal cache placement structure has  been entirely characterized in~\cite{Deng&Dong:TIT22}. As the delivery strategy in the MCCS is more complicated than that in the CCS, the cache placement optimization for the MCCS was only studied in~\cite{Jin&Cui:Arxiv2018} through numerical methods. The optimal cache placement structure for the MCCS  is still  unknown. Furthermore, in the existing coded caching schemes, since subfiles may be of different sizes, zero-padding is commonly used in the coded messages to simplify the delivery~\cite{Daniel&Yu:TIT19,Jin&Cui:Arxiv2017,Saberali&Lampe:TIT20,Deng&Dong:TIT22,Jin&Cui:Arxiv2018}.  However, the impact of zero-padding on the  coded delivery (for both the CCS and the MCCS) has, to our best knowledge, never been studied or known. 

\subsection{ Contributions} 

Our main goal in this paper is to characterize the memory-rate tradeoff for caching with uncoded placement under  nonuniform file popularity.
Later, we also extend our study to include nonuniform file sizes.
Our approach is to first formulate the cache placement optimization problem for the MCCS to minimize the average rate. We restrict the cache placement to the class of  popularity-first placements, which simplifies the optimization problem and has also been numerically shown to be optimal for the MCCS~\cite{Jin&Cui:Arxiv2018}. We then develop two information-theoretic lower bounds on the average rate for caching with  uncoded placement, one  for general cache placements and the other restricted to the popularity-first cache placements.
{By connecting and comparing the average rate of the optimized MCCS to the lower bounds, we characterize the memory-rate tradeoff under  nonuniform file popularity.} 

We partition the file request scenarios     into three regions to analyze, depending on the number of users $K$. For $K=2$ users, we prove that  the two lower bounds are identical. Furthermore, we show that the optimized MCCS achieves the lower bounds and is an optimal caching scheme under uncoded placement.
For $K>2$  users with distinct file requests, we show that the optimized MCCS is an optimal caching scheme under  popularity-first placement. For the above   two regions, our results for the MCCS also lead to the following two implications:  1) the popularity-first placement is optimal for the MCCS, and 2) zero-padding used in coded delivery incurs no loss of optimality. Finally, in the third region of $K>2$ users with redundant file requests,  we show that there may exist a  gap between the optimized MCCS and the lower bounds. Through analysis, we attribute the cause of this possible loss to  zero-padding used in the coded delivery  and quantify the loss.
{Even though the optimality of the MCCS in this region is uncertain in general, we provide some special cases  where  the MCCS is still shown to be optimal. Our numerical results further show that the loss  only exists  in some limited cases and is very small  in general.}

The region of $K>2$  users with distinct file requests also allows us to connect  the MCCS and the CCS under nonuniform file popularity. This enables us  to  further analyze the performance of the CCS that is otherwise  unknown in the literature. We show that 1) the use of zero-padding in the CCS~\cite{Daniel&Yu:TIT19,Jin&Cui:Arxiv2017,Saberali&Lampe:TIT20,Deng&Dong:TIT22} incurs no loss of optimality; and 2) for distinct file requests, the  optimized CCS~\cite{Daniel&Yu:TIT19,Jin&Cui:Arxiv2017,Saberali&Lampe:TIT20,Deng&Dong:TIT22} is an optimal scheme for caching under popularity-first placement.
  
With the understanding of the optimality of the MCCS, we next  characterize the optimal cache placement structure for the MCCS. By analyzing the cache placement optimization problem, we show that the possible structures for the optimal cache placement for the MCCS inherits that of the CCS obtained in~\cite{Deng&Dong:TIT22}.
Specifically, regardless of file popularity distribution,  there are at most three file groups in the optimal placement, where files in each group have the same placement. The optimal placement solution under each possible file grouping structure is obtained in closed-form.
The final optimal cache placement solution
is obtained by an efficient simple algorithm, which only requires computing a set of candidate solutions in closed-form   in parallel. {The  obtained optimal cache placement solution allows us to quantitatively evaluate the gap between  the optimized MCCS with the lower bounds in the region of $K>2$ users with redundant file requests to understand the exact memory-rate tradeoff.} The optimal cache placement solution provided by our algorithm is verified through simulations.
Note that
although the MCCS and the CCS share the same set of candidate placement structures, the optimal placement can still be different for the two  schemes, as a result of different coded delivery strategies. Numerical results  demonstrate this difference in the cache placement and  the performance gap between the optimized MCCS and the optimized CCS.

Lastly, we extend our study to the more general case where file popularity and sizes are both nonuniform.
We  formulate  the cache placement optimization problem for the MCCS and propose an information-theoretical lower bound on the average rate. We show that for $K=2$ users, the optimized MCCS achieves the proposed lower bound and thus characterizes the exact memory-rate tradeoff for caching with uncoded placement. For other cases, numerical results again show that the gap between the optimized MCCS and the lower bound is very small in general.  

\subsection{Related Works}

With a surge of interest in caching, there are many recent works study caching with uncoded placement. For uniform file popularity and sizes, the exact memory-rate tradeoff has been fully characterized for both  peak rate~\cite{Wan2016On1,Wan2016On2,Yu&Maddah-Ali:TIT2018} and average rate~\cite{Yu&Maddah-Ali:TIT2018}, which is achieved by the CCS and the MCCS, respectively. Beyond uncoded placement, the average rate of the optimized MCCS was shown to be at most a factor of two away from the optimal caching with any placement considered~\cite{Yu&Maddah19IT}.

When heterogeneity exists in the system, the characterization of the memory-rate tradeoff 
is generally an open problem.
For nonuniform file popularity, the cache placement problem  was studied for the CCS~\cite{Niesen&Maddah-Ali:TIT2017,Ji&Order:TIT17,hachem2017coded,Zhang&Coded:TIT18,Daniel&Yu:TIT19,Jin&Cui:Arxiv2017,Saberali&Lampe:TIT20,Deng&Dong:TIT22} and the MCCS~\cite{Jin&Cui:Arxiv2018} to minimize the achievable delivery rate.
For the CCS, to simplify the placement problem amid nonuniformity, suboptimal   file-grouping-based  cache placement strategies were proposed in~\cite{Niesen&Maddah-Ali:TIT2017,Ji&Order:TIT17,hachem2017coded,Zhang&Coded:TIT18}. They are shown to achieve an average rate that is a constant factor away from the lower bound for caching with any placement. Nonetheless, the gap is generally still large for practical consideration. Several works  used the optimization approach to study the cache placement problem \cite{Daniel&Yu:TIT19,Jin&Cui:Arxiv2017,Saberali&Lampe:TIT20}, either obtaining certain properties or devising numerical methods to solve the problem. The optimal placement structure has been completely characterized in~\cite{Deng&Dong:TIT22}, which shows inherit file grouping structure with at most three groups. For the MCCS, the complication in the improved delivery strategy adds challenges to the analysis, and the  cache placement problem was studied only in~\cite{Jin&Cui:Arxiv2018}. However,  the problem was numerically solved in that work, which cannot provide insight into the optimal cache placement structure. 

 Note that none of the above works~\cite{Niesen&Maddah-Ali:TIT2017,Ji&Order:TIT17,hachem2017coded,Zhang&Coded:TIT18,Daniel&Yu:TIT19,Jin&Cui:Arxiv2017,Jin&Cui:Arxiv2018,Saberali&Lampe:TIT20,Deng&Dong:TIT22}\ provided any lower bound for caching with uncoded placement to characterize the memory-rate tradeoff. The gap between the achievable rate of either the CCS or the MCCS and the optimal caching with uncoded placement remains unknown {under nonuniform file popularity and size}. Most recently, the exact memory-rate tradeoff under uncoded placement  for the case of two files~was characterized \cite{Sahraei2019TheOptimal}.
However, the caching scheme proposed in~\cite{Sahraei2019TheOptimal} is only designed for two files, which is not extendable to general scenarios. 

When files only have nonuniform  sizes, the CCS has again been shown to achieve  a peak rate a constant factor away from the lower bound for caching with any placement\cite{Zhang&Lin15Coded,Zhang&Lin19Closing}, where the gap may be large for practical concerns.
A limited number of recent works also considered joint nonuniformity in  cache size, file popularity and size \cite{Daniel&Yu:TIT19,Chang&Wang19Coded}. The cache placement optimization for the CCS was considered in~\cite{Daniel&Yu:TIT19}, where simplification methods were developed for the optimization problem with well-performed numerical solutions. However, ~\cite{Daniel&Yu:TIT19} focused on the optimization framework for the CCS, but did not address the optimality of the optimized CCS as compared to any information-theoretic lower bound.
In~\cite{Chang&Wang19Coded},  the memory-rate tradeoff   has been characterized under general placement, in the case of full nonuniformity in cache size, file popularity and size, but only for  a system of two users and two files, where a caching scheme was proposed to achieve the lower bound.
Except for these recent studies, the MCCS has never been explored for files with nonuniformity in both popularity and size.

Besides the above-mentioned works, coded caching schemes and the memory-rate tradeoff for caching have also been investigated in various systems or network configurations, including heterogeneous user profiles  \cite{Chang&Wang2019ISIT,Wang&Peleato19ISIT,Chang&Wang2020ICC,Zhang&Peleato2020ICC}, nonuniform cache sizes~\cite{Cao2019TCOM,Ibrahim2019Coded}, correlated files~\cite{Hassanzadeh2020TIT}, decentralized placement for nonuniform  file popularity, file size, and cache size~\cite{Wang2019Optimization}, heterogeneous distortion\cite{Yang18TIT,Hassanzadeh20TWCOM}, multi-antenna transmission and shared caches\cite{Parrinello2020Fundamental}.

\subsection{Organization and Notations}\label{subsec:notation}
The rest of the paper is organized as follows. The system model is presented in Section~\ref{sec:model}. In Section~\ref{sec:prob}, we formulate the cache placement optimization problem for the MCCS under nonuniform file popularity. In Section~\ref{sec:lb}, we propose two lower bounds  for caching with uncoded placement and discuss the relation of the two bounds. In Section~\ref{sec:exact}, we characterize the memory-rate tradeoff for caching by comparing the optimized MCCS with the two lower bounds and identify the optimality of the MCCS in certain regions.
In Section~\ref{sec:MCCS_placement}, we derive the optimal cache placement structure for the MCCS under nonuniform file popularity. In Section~\ref{sec:non_length_popu}, we extend our study to  files with nonuniform  popularity and size to characterize the memory-rate tradeoff. Numerical results are provided in Section~\ref{sec:num}, followed by  the conclusion  in Section~\ref{sec:conclusion}.

\emph{Notations:} The cardinality of set $\Sc$ is denoted by $|\Sc|$, and   the index set for $\Sc$ is defined by $\Ic_{|\Sc|}=\{1,\ldots,|\Sc|\}$. The size of file $W$ is denoted by $|W|$.  The bitwise "XOR" operation between two subfiles is denoted by
 $\oplus$. Notations $\lfloor\cdot\rfloor$ and $\lceil\cdot\rceil$ denote the floor and ceiling functions, respectively. Notation  $\abf \succcurlyeq {\bf 0}$ means vector $\abf$ is element-wise  non-negative. 
Also, we extend the definition of ${K \choose l}$ and define ${K \choose l}=0$,  for  $l<0$ or $l>K$.

\allowdisplaybreaks
\section{ System Model}\label{sec:model}
  \begin{figure}[t]
    \centering
    \includegraphics[scale=0.35]{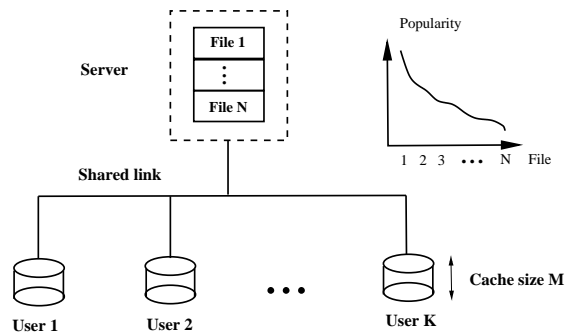}
    \renewcommand{\figurename}{Fig.}
    \caption{A cache-aided system with end users equipped with a local cache connecting to the server via a shared link.} \label{fig:sys_mod}
  \end{figure}

We consider a cache-aided transmission system with a server connecting to $K$ cache-equipped users over a shared error-free link, as shown in Fig. \ref{fig:sys_mod}.
The server has a database consisting of $N$ files $\{W_1,\ldots,W_N\}$. Each file $W_n$ is of size $F$ bits and has  probability  $p_n$ of being requested. The popularity distribution of  the entire $N$ files is denoted by $\pbf\triangleq[p_1,\ldots,p_N]^T$, with $\sum_{n=1}^{N}p_n=1$.
The files are indexed according to the decreasing order of their popularities as $p_1\geq p_2\geq \cdots\geq p_N$.
Each user $k$  has a local cache, whose size in the unit of file is  $M$ files, representing a cache capacity of $MF$ bits.   The size $M$ can be an arbitrary real number within interval $[0,N]$.
We denote   $\Nc\triangleq\{1,\ldots,N\}$  and $\Kc\triangleq\{1,\ldots,K\}$.

The coded caching operates in  the cache placement phase and the content delivery phase.
In the cache placement phase, under a cache placement scheme, a portion of uncoded file contents are placed in each user's\ local cache.
Assume each user $k$ independently requests a file with index $d_k$  from the server.
Let  $\dbf\triangleq[d_1,\ldots,d_K]^T$ denote the demand vector of $K$ users. In the content delivery phase, based on the demand vector  $\dbf$ and the cached contents at users, the server generates coded messages containing those uncached portions of requested files and transmits these coded messages to the users.
Upon receiving the coded messages, each user $k$  reconstructs its requested file $W_{d_k}$  from the received coded messages and its cached content. With a valid coded caching scheme,  each user $k$ is able to reconstruct its requested file for any demand vector $\dbf\in\Nc^K$ over an error-free link.

\section{Cache Placement for Rate Minimization} \label{sec:prob}
For any coded caching scheme, cache placement is a key design issue, which needs to be optimized to  minimize the delivery rate. The MCCS is a coded caching scheme recently proposed \cite{Yu&Maddah-Ali:TIT2018}, where the delivery strategy is improved over the original CCS \cite{Maddah-Ali&Niesen:TIT2014}   to  reduce the delivery rate further. In this section, we formulate  the rate minimization problem for the MCCS under the cache placement optimization.

\subsection{Cache Placement} \label{III.A}
The cache placement construction for the MCCS is based on file partitioning. For $K$ users, there are total $2^K$ user subsets in $\Kc$, with subset sizes ranging from $0$ to $K$ (including the empty set). Grouping the user subsets based on their sizes, we form a cache subgroup that contains all user subsets of size $l$, defined as $\Ac^l\triangleq\{\Sc: |\Sc|=l,\ \Sc\subseteq \Kc\}$ with $|\Ac^l|=\binom{K}{l}$,  for $l=0,\ldots,K$. Partition each file $W_n$  into $2^K$ non-overlapping subfiles. Each subfile is for a unique user subset $\Sc\subseteq\Kc $, denoted by
 $W_{n,\Sc}$, and it is stored at the local cache of each user in subset $\Sc$. It is possible that $W_{n,\Sc}=\emptyset$ for a given $\Sc$, and also for $\Sc=\emptyset$,  subfile $W_{n,\emptyset}$  is only kept   in the server and not stored in any user's cache. A caching scheme specifies how files are partitioned for storage. Regardless of the scheme used, each  file should be able to  be reconstructed  by combining all its subfiles. Thus, we have
\begin{align}\label{cons:file_partition_0}
\sum_{l=0}^{K}\sum_{\Sc\in \Ac^l}|W_{n,\Sc}|=F,\quad  n \in \Nc.
\end{align}

There are $2^K$ subfile sizes to be determined for each file. To reduce  the number of variables and simplify the cache placement  problem for its tractability, the following condition is imposed:
\begin{assumption}
{For any  file $W_n$, the size of its subfile $W_{n,\Sc}$ is the same for  any $\Sc$ of the same size, \ie $|W_{n,\Sc_1}|=|W_{n,\Sc_2}|$, for $\forall~\Sc_1,\Sc_2 \in \Ac^l$, $l=0,\cdots,K$.}
\end{assumption}
The above condition is in fact proven to be the property of the optimal cache placement for the CCS~\cite{Jin&Cui:Arxiv2017}. For the MCCS, although it is more difficult to prove analytically, it is numerically verified in \cite{Jin&Cui:Arxiv2018} that imposing this condition results in no loss of optimality.\footnote{In Section \ref{sec:exact}, we are able to prove that imposing Condition C1) does not incur loss of optimality in some specific cases. } 
As a result, the subfiles of file $W_n$  are grouped  into file subgroups according to user subset size $l$, each denoted  by $\Wc^l_n=\{W_{n,\Sc}: \Sc \in \Ac^l\}$, for $l=0,\ldots,K$. Note that there are ${K \choose l}$ subfiles of the same size in  $\Wc^l_n$, and there are  $K+1$ file subgroups. Following this, let $a_{n,l}$ denote  the  size of subfiles in $\Wc^l_n$ as a fraction of  file $W_n$ of size $F$ bits,  \ie $a_{n,l} \triangleq |W_{n,\Sc}|/F$, for $\forall \Sc \in \Ac^l$,  $l=0,\ldots,K$,   $n\in \Nc$.  In particular, $a_{n,0}$ represents the fraction of file $W_n$  that is not stored at any user's cache but only remains in the server. Then, the file partition constraint \eqref{cons:file_partition_0} is simplified to
\begin{align}
  \sum_{l=0}^{K}{K \choose l}a_{n,l}=1, \ n \in \Nc.\label{Constraint1.1}
\end{align}

Recall that each subfile is intended for a unique user subset. For the cache placement, user $k$  stores all the subfiles in $\Wc_n^l$ that are  intended for it, \ie $\{W_{n,\Sc}: k\in \Sc \text{~and~} \Sc \in \Ac^{l+1}\}\subseteq \Wc_n^l$, for  $l=0,\ldots,K-1$.
Note that in each cache subgroup $\Ac^l$, there are  ${K-1 \choose l-1}$  different user subsets containing the same user $k$. Thus,  there are  $\sum_{l=1}^{K}{K-1 \choose l-1}$ subfiles in each file $W_n$ that a user can  store in its local cache. This means  that, in total, a fraction  $\sum_{l=1}^{K}{K-1 \choose l-1}a_{n,l}$  of file $W_n$ is cached by a user. With  cache size $M$ at each user, we have the following cache  constraint
\begin{align}
\sum_{n=1}^{N}\sum_{l=1}^{K}{K-1 \choose l-1}a_{n,l} \leq M.\label{Constraint2}
\end{align}

For nonuniform file popularity, even with Condition C1, the cache placement is still a complicated problem. To further  simplify the cache placement problem and the average rate expression, we consider the popularity-first cache placement approach described below.

\vspace*{.5em}
\noindent \textbf{\textit{Popularity-first cache placement}}: 
A popularity-first
cache placement is to allocate more cache memory to a more
popular file: With file popularity $p_1 \ge \cdots \ge p_N$, the cached subfiles satisfies
\begin{align}\label{equ:popu_fir}
a_{n,l}\ge a_{n+1,l},\quad l\in\Kc,~n\in\Nc\backslash\{N\}.
\end{align}
\begin{Remark}\label{remark1}
The popularity-first  cache placement approach has been used for both the CCS\cite{Daniel&Yu:TIT19,Jin&Cui:Arxiv2017} and the MCCS\cite{Jin&Cui:Arxiv2018} to simplify the cache placement problem. For the CCS, the popularity-first placement has been proven to be the property of the optimal cache placement\cite{Jin&Cui:Arxiv2017}. For the MCCS, the same is difficult to prove analytically, but the optimality of the popularity-first placement  has been verified  numerically   \cite{Jin&Cui:Arxiv2018}. In Section \ref{sec:exact}, we will prove the optimality of popularity-first placement in some cases.
\end{Remark}

\subsection{Content Delivery}\label{subsec:delivery}
In the content delivery phase, the server multicasts coded messages to  different user subsets. Each coded message corresponds to a user subset $\Sc$, formed by the bitwise XOR operation  of subfiles as
\begin{align} \label{C_S}
C_\Sc\triangleq\bigoplus_{k \in \Sc} \! W_{d_k,\Sc\backslash\{k\}}.
\end{align}

In the original  CCS \cite{Maddah-Ali&Niesen:TIT2014}, the server simply delivers the coded message  formed by each user subset, for any demand vector  $\dbf$. However, under random demands, multiple users may request the same (popular) file, causing redundant coded messages transmitted  separately multiple times.
To address this, in the MCCS \cite{Yu&Maddah-Ali:TIT2018}, a modified coded delivery strategy is proposed to remove this   redundancy and reduce the average delivery rate further. Let $\widetilde{N}(\dbf)$  denote the number of distinct file requests for demand vector $\dbf$, where $\widetilde{N}(\dbf)\le K$. To describe the delivery strategy in the MCCS, we  provide the following four definitions:
\begin{myDef}[Leader group]\label{defLeader}
The leader {group} $\Uc$ is a user subset of size   $|\Uc|=\widetilde{N}(\dbf)$, with the users in  $\Uc$ having exactly $\widetilde{N}(\dbf)$ distinct file requests.
\end{myDef}
\begin{myDef}[Redundant group]\label{defRedundant}
Given the leader group $\Uc$, any user subset $\Sc \subseteq \Kc$ with $\Sc \cap \Uc=\emptyset$ is called a redundant {group};  otherwise, $\Sc$ is a non-redundant {group}.
\end{myDef}

\begin{myDef}[Redundant request]\label{defRedundant_req}
A file request $d_k$ by  any user $k$ in the redundant group $\Sc$ is a redundant request.
\end{myDef}

\begin{myDef}[Redundant message]
Any coded message  $C_\Sc$ corresponding to a  redundant group $\Sc$ is a redundant message; otherwise, it is a non-redundant message.
\end{myDef}

Based on the above definitions, any user subset is either a redundant group or a non-redundant group. In the MCCS, only the non-redundant messages are  multicasted  to both non-redundant and redundant groups.  For nonuniform file popularity,
file partitioning may be different for different files, leading to different subfile sizes. In formulating the coded message $C_\Sc$ in \eqref{C_S}, the following technique is commonly used for  the subfiles of different lengths in the XOR operation: 
\vspace*{.5em}

\noindent \textbf{\textit{Zero-padding}}: With different subfile sizes, subfiles in  coded message $C_\Sc$ are zero-padded to the size of the
largest subfile in $C_\Sc$ for transmission. 
\vspace*{.5em}

Consider  zero-padding for the coded message $C_\Sc$ for subgroup $\Sc$. The size of $C_\Sc$ in \eqref{C_S} is given by
\begin{align} \label{equ:zero_pad}
|C_\Sc|=\max_{k\in\Sc}a_{d_k,l},\quad \Sc\in\Ac^{l+1}, \; l=0,\ldots,K-1.
\end{align}

\begin{Remark}\label{remark2}
Zero-padding is a technique commonly used to form coded messages in the existing works~{\cite{Niesen&Maddah-Ali:TIT2015,Niesen&Maddah-Ali:TIT2017,Ji&Order:TIT17,Zhang&Coded:TIT18,Daniel&Yu:TIT19,Jin&Cui:Arxiv2017,Jin&Cui:Arxiv2018,Deng&Dong:TIT22,Saberali&Lampe:TIT20}.} {In some proposed delivery schemes, zero-padding may be limited to coding subfiles within a file group, such as in  \cite{Niesen&Maddah-Ali:TIT2017,Ji&Order:TIT17,Zhang&Coded:TIT18}. Zero-padding considered in \eqref{equ:zero_pad} for $C_{\Sc}$ is general for  any files requested by a user subset $\Sc$.} Despite being a common technique, the impact of zero-padding on  coded caching  has not been analyzed and is unknown.   Intuitively, zero-padding   introduces extra waste bits that may degrade the performance. In Section~\ref{sec:exact}, we will  provide   our findings and insight on this issue.
\end{Remark}

\subsection{Cache Placement Optimization}

Let $\abf_n\triangleq[a_{n,0},\ldots,a_{n,K}]^T$ denote the $(K+1)\times 1$ cache placement vector for file $W_n$, $n\in\Nc$, and let $\abf\triangleq[\abf_1^T,\cdots,\abf_N^T]^T$ represent the entire placement for $N$ files.
For demand vector $\dbf$, the delivery rate is the total size of  the non-redundant messages, given by
\begin{align}\label{CodedMsgPad}
R_{\text{MCCS}}(\dbf;\abf)&=\!\!\sum_{\Sc\subseteq\Kc,\Sc \cap \Uc\neq\emptyset}\!\!|C_\Sc|=\!\!\sum_{\Sc\subseteq\Kc,\Sc \cap \Uc\neq\emptyset}\max_{k\in\Sc}a_{d_k,l}.
\end{align}
 The average delivery rate $\bar R_{\text{MCCS}}$ is given by
\begin{align}\label{avgR_MCCS}
 \bar{R}_{\text{MCCS}}(\abf)=\mathbb{E}_\dbf\left[R_{\text{MCCS}}(\dbf;\abf)\right]
=\mathbb{E}_\dbf\Bigg[\sum_{\Sc\subseteq\Kc,\Sc \cap \Uc\neq\emptyset}\max_{k\in\Sc}a_{d_k,l}\!\Bigg]
\end{align}
where $\mathbb{E}_\dbf[\cdot]$ is taken with respect to $\dbf$.

From \eqref{equ:popu_fir}, we define the set of all popularity-first placements by $\Qc\triangleq\{\abf:\; a_{n,l}\ge a_{n+1,l}, \; l\in\Kc,  n\in\Nc\backslash\{N\}\}$. {Here, we assume that $F$ is large enough such that $a_{n,l}F \in\mathbb{Z}$}. To obtain the minimum average rate for the MCCS, we  optimize the  cache placement    $\abf\in\Qc$ to minimize $\bar{R}_\text{MCCS}$, given by\footnote{The assumption of $F$ being large is common in the existing works \cite{Niesen&Maddah-Ali:TIT2017,hachem2017coded,Ji&Order:TIT17,Zhang&Coded:TIT18,Jin&Cui:Arxiv2017,Saberali&Lampe:TIT20,Deng&Dong:TIT22,Cheng&Li17Optimal,Zhang&Lin15Coded,Zhang&Lin19Closing,Daniel&Yu:TIT19,Sahraei2019TheOptimal,Jin&Cui:Arxiv2018}. In practice, the file size  typically  exceeds $1$ kbit or $1$ Mbit, for which the formulation in {\bf P0} becomes accurate, \ie  $a_{n,l}F\in\mathbb{Z}$. Even in the case of very small $F$, {\bf P0}  can be viewed as the relaxed version of the original problem with $a_{n,l}F\in\mathbb{Z}$. We can round the optimal solution $a_{n,l}F$ to {\bf P0} to the nearest integer solution. The same discussion applies to {\bf P1} and {\bf P2} in Section~\ref{sec:lb} as well.}
\begin{align}
\textrm{\bf P0}: \;\min_{\abf\in\Qc}&\;\;  \bar{R}_{\text{MCCS}}(\abf) \quad\\
 \textrm{s.t.} &\;\;
\eqref{Constraint1.1},\eqref{Constraint2}, \; \text{and~}\nn\\
&\;\; \abf_{n}\succcurlyeq\mathbf{0},\; n\in\Nc. \label{Constraint_gt0}
\end{align}

Note that in {\bf P0}, we restricted the cache placement optimization within the set of popularity-first placements, for the reason discussed  in Remark~\ref{remark1}. In the following,  we first focus on analyzing  how optimal the MCCS in  {\bf P0} is under nonuniform file popularity, by comparing it with  the lower bounds we develop for caching with uncoded placement. Then, in  Section \ref{sec:MCCS_placement}, we describe the optimal cache placement solution to   {\bf P0} and its inherent structure.

\section{Converse Bound for Uncoded Placement}\label{sec:lb}
In this section, we  first introduce a lower bound on the average rate for any caching with  uncoded placement. Then, we develop a popularity-first-based lower bound by restricting the uncoded placement to the  set of popularity-first placements.

Let $\Dc$ denote the set of the  distinct file indices  in demand vector $\dbf$, \ie  $\Dc = \Unique(\dbf)\subseteq \Nc$, where $\Unique(\dbf)$ is to extract the unique elements in $\dbf$.
{Recall the definition of index set $\Ic_{|\Dc|}$ for $\Dc$ is given in Section~\ref{subsec:notation}.} The following lemma gives a lower bound on the average rate under any uncoded placement.

\begin{lemma}\label{lemma_bnd_2}
For the caching problem described in Section~\ref{sec:model}, the following optimization problem provides a lower bound on the average rate for caching with uncoded placement:
\begin{align}
\textrm{\bf P1:}\;\; \min_{\abf}\; \bar R_\text{lb}(\abf)&\triangleq \sum_{{\Dc}\subseteq\Nc} \sum_{\dbf\in\Tc({\Dc})}\prod_{k=1}^{K}p_{d_k} R_\text{lb}({\Dc};\abf) \label{equ:P1_obj}\\ \quad
      \textrm{s.t.} &\quad \eqref{Constraint1.1},\eqref{Constraint2}, \text{~and~} \eqref{Constraint_gt0}\nn
\end{align}
where $\Tc({\Dc})\triangleq \{\dbf: \Unique(\dbf)=\Dc, \ \dbf\in \Nc^K\}$, and $R_\text{lb}({\Dc};\abf)$ is the  lower bound for the distinct file set ${\Dc}$ with the placement vectors $\{\abf_n, n\in{\Dc}\}$, given by
\begin{align}\label{R_lblb}
R_\text{lb}(\Dc;\abf)\triangleq\max_{\pi:\Ic_{|\!\Dc\!|}\rightarrow\Dc} \sum_{l=0}^{K-1}\sum_{i=1}^{|\Dc|}\binom{K-i}{l}a_{\pi(i),l}
\end{align}
where $\pi\!:\Ic_{|\!\Dc|}\!\to\!\Dc$ is any bijective map from $\Ic_{|\!\Dc|}$ to     ${\Dc}$.
\end{lemma}
\IEEEproof
See Appendix~\ref{Proof:lemma_bnd_2}.
 \endIEEEproof

Note that {\bf P1} is a min-max problem. It can be  cast  in its epigraph form by moving \eqref{R_lblb} to the constraints. The resulting equivalent problem is a linear program (LP), which can be solved by  standard LP  solvers.
\begin{Remark}
Recall that to simplify the cache placement problem, the cache placement vector $\abf$ for the MCCS is formed under Condition C1. Here, we point out that although   {\bf P1} is w.r.t. the same cache placement vector $\abf$,  the formation of $R_\text{lb}(\Dc;\abf)$ in \eqref{R_lblb} does not require us to impose Condition C1). This can be seen from the derivation  in the proof in Appendix~\ref{Proof:lemma_bnd_2}. Thus, the lower bound in {\bf P1} is not based on Condition C1, but rather the  placement $\abf$ that satisfies Condition C1 turns out to be optimal to {\bf P1}. 
\end{Remark}
\begin{Remark}
We point out a subtle yet key difference between our lower bound from the existing ones. In characterizing the caching limit, the challenge in the existing works lies in analyzing the gap between the lower bound and the specific caching scheme used. For example,  it is difficult to see the connection between the lower bound obtained in \cite{Sahraei2019TheOptimal} and  any achievable scheme for comparison. Although the derivation of the lower bound in Lemma~\ref{lemma_bnd_2} is based on the approach in \cite{Sahraei2019TheOptimal},    as shown in Appendix \ref{Proof:lemma_bnd_2}, we are able to transform the lower bound  into an equivalent form, which  satisfies Condition C1. This transformation is the result of the in-depth understanding of the cache placement optimization formulation for the MCCS. It paves the way for  connecting the lower bound in {\bf P1} (and the lower bound in {\bf P2} below) to the optimized MCCS and analyze the gap. 
\end{Remark}

Given that  the popularity-first placement approach has been considered  in the existing works under nonuniform file popularity,  we also  develop a popularity-first-based lower bound, by imposing  the  popularity-first placement constraint  \eqref{equ:popu_fir} to the placement vector $\abf$ in {\bf P1}, as shown below.

\begin{lemma}\label{lemma:lbpf}
(Popularity-first-based lower bound) For the caching problem described in Section~\ref{sec:model}, the following  optimization problem provides a lower bound on the average rate for caching under  popularity-first  cache placement:
\begin{align}
\textrm{\bf P2:}\; \min_{\abf\in\Qc}\; \bar R_\text{lb}(\abf)&\triangleq \sum_{{\Dc}\subseteq\Nc} \sum_{\dbf\in\Tc({\Dc})}\prod_{k=1}^{K}p_{d_k} R_\text{lb}({\Dc};\abf)\label{equ:converse_obj_pf}\\
\quad      \textrm{s.t.} &\quad \eqref{Constraint1.1},\eqref{Constraint2}, \ \text{and}\  \eqref{Constraint_gt0}\nn
\end{align}
 where $R_\text{lb}({\Dc};\abf)$, for $\abf\in\Qc$, is given by
\begin{align}
 R_\text{lb}(\Dc;\abf)\triangleq \sum_{l=0}^{K-1}\sum_{i=1}^{|\Dc|}\binom{K-i}{l}a_{\phi(i),l},\quad \abf\in\Qc \label{equ:bnd_5}
\end{align}
where $\phi\!:\Ic_{|\!\Dc|}\!\to\!\Dc$ is the bijective map in the decreasing order of file popularity, \ie $p_{\phi(1)} \ge \cdots \ge p_{\phi(|\!\Dc|)}$.
\end{lemma}
\IEEEproof
See Appendix \ref{Proof:lemma_lbpf}.
\endIEEEproof

Note that for $R_\text{lb}(\Dc;\abf)$  in \eqref{R_lblb}, by restricting $\abf \in \Qc$, we can remove the max operation and simplify the expression to \eqref{equ:bnd_5}. Due to the restriction $\abf\in\Qc$ in {\bf P2}, the lower bound given by {\bf P1}  is also a lower bound for {\bf P2}. In the following theorem, we show that for $K=2$ users,  there is no loss of optimality by restricting to popularity-first placements.
\begin{theorem}\label{thm:lbequ}
For $K=2$, {\bf P1} and {\bf P2} are equivalent.
\end{theorem}
\IEEEproof
See Appendix \ref{Proof:thm:lbequ}
\endIEEEproof

Theorem \ref{thm:lbequ} indicates that, with nonuniform file popularity, for  $K=2$ users,   the popularity-first placement is {an optimal solution to {\bf P1}}, regardless of the values of $N$, $\pbf$, and $M$. In other words, the general lower bound  for any caching with uncoded placement  given by {\bf P1} is attained by some $\abf \in \Qc$. {We will use this result  in Section~\ref{sec:two-user} to further show that the MCCS under the popularity-first placement is an optimal caching scheme under  uncoded placement for $K=2$ users. For $K>2$ users,  except for some special cases, the same conclusion turns out to be challenging to prove in general. Although  the same conclusion cannot be shown analytically,  we provide numerical results in Section~\ref{sec:num} to show that the two lower bounds by {\bf P1} and {\bf P2} are  equal in general, and  a very small gap is observed only in some limited cases}.

\section{Memory-Rate Tradeoff Characterization  }\label{sec:exact}
In this section, we compare the average rate of the optimized MCCS in {\bf P0} and that of the popularity-first-based lower bound in {\bf P2} to show the tightness of the bound and  how optimal the MCCS is.
Consider $N$ files with arbitrary popularity distribution $\pbf$ and local cache size $M$.  We partition  the number of users  $K$ and their file requests in three possible regions to compare  {\bf P0} and  {\bf P2}:\\[.2em]
\emph{Region 1}: $K=2$;\\
\emph{Region 2}: $K> 2$, $\tilde N(\dbf)=K$ (no  redundant file requests);\\ \emph{Region 3}: $K>2$, $\tilde N(\dbf)<K$ (with redundant file requests).\\[-1em]

Note that  Region 2 is possible only when  $K\le N$, and Region 3 is when there are multiple users  requesting  the same file.
We summarize our results below:
\begin{itemize}
  \item For both Regions 1 and 2, we prove that the popularity-first-based lower bound in {\bf P2} is tight, \ie   the optimized MCCS in {\bf P0} attains this lower bound. In particular, in Region 1,   by Theorem~\ref{thm:lbequ}, the result  further indicates the  MCCS is an optimal caching scheme (with the optimized popularity-first placement) for any caching with uncoded placement. Also,  the tight bound reveals that there is no loss of optimality by  zero-padding  in coded messages in the MCCS in  Regions 1 and 2.
\item For Region 3, we show that there may be a performance gap between the optimized MCCS and the popularity-first-based lower bound in {\bf P2}.     The  loss is due to zero-padding in the coded messages in the delivery phase, which we will discuss and quantify. Nonetheless, {we provide some special cases or a condition for which  {\bf P0} and  {\bf P2} are equivalent. Also}, our numerical evaluation of the optimized MCCS and the lower bound in Section~\ref{sec:num} show that, in general, the loss only appears in some limited  scenarios and is very small.
\end{itemize}

{We describe  the above results in detail in the following subsections. Note that the difference between {\bf P0} and {\bf P2} is only  in the expression of the average rate objective function. Thus, we focus on comparing $\bar R_\text{MCCS}(\abf)$ and $\bar R_\text{lb}(\abf)$,  for  $\abf\in\Qc$. Note that for {\bf P0},  $R_\text{MCCS}(\dbf;\abf)$ in \eqref{CodedMsgPad} is a function of  demand vector $\dbf$, while   for {\bf P2},  $R_\text{lb}(\Dc; \abf)$ in \eqref{equ:bnd_5} is a function of distinct file subset $\Dc$. Furthermore, $R_\text{MCCS}(\dbf;\abf)$ in \eqref{CodedMsgPad} contains the max operation, while $R_\text{lb}(\Dc; \abf)$ in \eqref{equ:bnd_5} does not.  Thus, it is challenging  to connect and compare the two objective functions, especially given the complicated delivery scheme in the MCCS. Our first step is  to reformulate the expression of $R_\text{MCCS}(\dbf;\abf)$   such that we can represent the max operation in a way to connect the expression to the lower bound $R_\text{lb}(\Dc; \abf)$ in the later steps.  
}

\subsection{Expression of $\bar R_\text{MCCS}(\abf)$} \label{sec:achiev}
We first rewrite the expression of $\bar R_\text{MCCS}(\abf)$ in   \eqref{avgR_MCCS} for the MCCS.
Given  placement vector $\abf\in\Qc$, we rewrite   $R_{\text{MCCS}} (\dbf;\abf)$  in   \eqref{CodedMsgPad} for demand vector $\dbf$  as
\begin{align}
R_{\text{MCCS}} (\dbf;\abf) =\sum_{l=0}^{K-1}\sum_{\Sc\in\Ac^{l+1}, \Sc \cap \Uc\neq\emptyset}\!\!\!\max_{k\in\Sc}a_{d_k,l} \label{equ:converse_achi_obj}
\end{align}
where we regroup the terms in $R_{\text{MCCS}} (\dbf;\abf)$ based on the size $|\Sc|$ of the non-redundant groups, and  $\Uc$ is the leader group for $\dbf$. Define $\psi\!:\Ic_{|\Uc|}\!\to\!\Uc$ as a bijective map from $\Ic_{|\Uc|}$ to the leader group $\Uc$, such that the requested (distinct) files by the users in   $\Uc$ are ordered in decreasing  popularity, \ie $p_{d_{\psi(1)}}\ge\ldots\ge p_{d_{\psi(|\Uc|)}}$. Recall that  $\phi\!:[|\Dc|]\!\to\!\Dc$ defined in Lemma~\ref{lemma:lbpf} maps the indices of distinct files with the same file popularity order. {Also,  from Definition~\ref{defLeader} of the leader group $\Uc$, we note that $|\Dc|=|\Uc|=\tilde{N}(\dbf)$, for  given $\dbf$.} Thus, by the relation of $\Dc$ and $\Uc$, the two mappings  $\psi(\cdot)$ and $\phi(\cdot)$ are based on the same file popularity order, and we have  $d_{\psi(i)}=\phi(i)$, $i=1,\ldots,\tilde N(\dbf)$. Also, since  $\abf\in\Qc$, we have
\begin{align}\label{a:psi_order}
a_{d_{\psi(1)},l}\geq\ldots\geq a_{d_{\psi(\tilde N(\dbf))},l}, \quad l\in \Kc.
\end{align}

 To evaluate the inner max operation  in \eqref{equ:converse_achi_obj}, we partition the coded messages into different categories according to the user subsets. Recall that cache subgroup $\Ac^{l+1}$ is the set of all  $\binom{K}{l+1}$ user subsets with size $|\Sc|=l+1$. Among these user subsets, there are $\binom{K-1}{l}$ subsets containing  user $\psi(1)$ in $\Uc$.
Let  $\bar{a}_{\psi(1),l}$ denote the size of the coded message corresponding to each of these $\binom{K-1}{l}$ subsets containing user $\psi(1)$.  From \eqref{a:psi_order}, we have
\begin{align}
\bar{a}_{\psi(1),l}^\sSc= \max_{k\in\Sc  } a_{d_k,l}, \; \text{for~} \psi(1)\in\Sc\cap \Uc, \Sc\in \Ac^{l+1}. \label{equ:ach_obj_L1}
\end{align}
Similarly, there are $\binom{K-2}{l}$ user subsets in $\Ac^{l+1}$ that contain user $\psi(2)$ but not $\psi(1)$ in $\Uc$. Denote the size of the coded message corresponding to each of these subsets as $\bar{a}_{\psi(2),l}$, then
 \begin{align}
 \bar{a}_{\psi(2),l}^\sSc=\max_{k\in\Sc}a_{d_k,l}, \; \text{for~}  \psi(1) \notin \Sc, \psi(2)\in  \Sc\cap \Uc, \Sc\in \Ac^{l+1}.  \label{equ:ach_obj_L2}
 \end{align}
Following the above,  in general, the number of user subsets in $\Ac^{l+1}$ that include $\psi(i)$ but not  $\psi(1),\ldots,\psi(i-1)$ is $\binom{K-i}{l}$. Let  $\bar{a}_{\psi(i),l}$ denote the size of the coded message corresponding to each of these subsets. Then, we have
\begin{align}\label{equ:ach_obj_L3}
\bar{a}_{\psi(i),l}^\sSc=\max_{k\in\Sc }a_{d_k,l}, \ \text{for~} \Sc\in \tilde{\Ac}^{l+1}_i
\end{align}
where $\tilde{\Ac}^{l+1}_i\triangleq \{\Sc \in \Ac^{l+1}\!\!:  \{\psi(1),\ldots,\psi(i-1)\}\cap \Sc = \emptyset,  \psi(i)\in\Sc \cap \Uc\}$, with $|\tilde{\Ac}^{l+1}_i|= \binom{K-i}{l}$.

Based on \eqref{equ:ach_obj_L3}, we can rewrite \eqref{equ:converse_achi_obj} as
\begin{align}\label{equ:achiev_conv_1}
R_{\text{MCCS}}(\dbf;\abf) = \sum_{l=0}^{K-1}\sum_{i=1}^{\tilde{N}(\dbf)}\sum_{\Sc \in \tilde{\Ac}^{l+1}_i}\bar{a}_{\psi(i),l}^\sSc.
\end{align}
Averaging $R_{\text{MCCS}}(\dbf;\abf)$ in \eqref{equ:achiev_conv_1} over $\dbf$,
 $\bar R_\text{MCCS}(\abf)$  in \eqref{avgR_MCCS}  can be rewritten as
 \begin{align}\label{equ:avg_R_MCCS}
\bar R_\text{MCCS}(\abf)=\sum_{\Dc\subseteq\Nc}\sum_{\dbf\in \Tc({\Dc})}\prod_{k=1}^{K}p_{d_k} R_{\text{MCCS}}(\dbf;\abf).
\end{align}
where $\Tc({\Dc})$ is defined below {\bf P1} in Lemma~\ref{lemma_bnd_2}. With the expression in \eqref{equ:avg_R_MCCS}, we now can directly compare the minimum average rate in {\bf P0} with {\bf P1} or {\bf P2}.

\subsection{Region 1: $K=2$}\label{sec:two-user}
In this region, We have the following result on the optimality of the MCCS.

\begin{theorem}\label{thm_K2}
For  a caching problem of   $N$ files with popularity distribution $\pbf$ and local cache size $M$, for $K=2$ users,  the minimum average rate of the optimized MCCS in {\bf P0} attains the lower bound given by {\bf P1}. Thus, the  MCCS is  optimal  for caching with uncoded placement.
\end{theorem}

\IEEEproof
We first show that   {\bf P0} and {\bf P2} are equivalent, \ie $\bar R_\text{MCCS}(\abf)=\bar R_\text{lb}(\abf)$, for $\abf\in\Qc$. Comparing the two expressions  in \eqref{equ:converse_obj_pf} and \eqref{equ:avg_R_MCCS}, we  only need to examine  $R_{\text{MCCS}}(\dbf;\abf)$ and $R_\text{lb}({\Dc};\abf)$. For $\Kc=\{1, 2\}$, we have $|\Dc|=\tilde N(\dbf)=1$ or $2$. We consider these two cases separately below.

\emph{Case 1) $\tilde{N}(\dbf)=1$}:
 In this case, two users request the same file. Based on the relation of  two mappings $\psi(\cdot)$ and $\phi(\cdot)$ discussed in Section~\ref{sec:achiev}, we have\  $\Dc=\{\phi(1)\}$, and $d_{\psi(1)}=d_{\psi(2)}=\phi(1)$. For $K=2$,  $R_\text{lb}({\Dc;\abf})$ in \eqref{equ:bnd_5}  is given by
 \begin{align}\label{equ:gap_K2_1}
 R_\text{lb}({\Dc;\abf})&=a_{\phi(1),0}+a_{\phi(1),1}.
 \end{align}
 For  $R_{\text{MCCS}}(\dbf;\abf)$ in \eqref{equ:achiev_conv_1},  based on  $\tilde{\Ac}^{l+1}_i$ defined below \eqref{equ:ach_obj_L3}, for $\tilde{N}(\dbf)=1$, we have
 $\tilde{\Ac}^{1}_1=\{\psi(1)\}$ and $\tilde{\Ac}^{2}_1=\{\psi(1),\psi(2)\}$. Thus, from \eqref{a:psi_order} and    $\bar{a}_{\psi(1),l}^\sSc$ in \eqref{equ:ach_obj_L1}, we have
\begin{align}\label{equ:gap_K2_0}
 R_{\text{MCCS}}(\dbf;\abf)&=\!\!\!\sum_{\Sc \in \tilde{\Ac}^1_1}\!\!\bar{a}_{\psi(1),0}^\sSc +\!\!\!\sum_{\Sc \in \tilde{\Ac}^2_1}\!\!\bar{a}_{\psi(1),1}\nn\\
&=a_{d_{\psi(1)},0}+a_{d_{\psi(1)},1}.
  \end{align}
Since $d_{\psi(1)}={\phi(1)}$, from \eqref{equ:gap_K2_1} and \eqref{equ:gap_K2_0}, we have
\begin{align}\label{equ:gap_K2_11}
 \!\!\!\!R_{\text{MCCS}}(\dbf;\abf)=a_{\phi(1),0}\!+\!a_{\phi(1),1}\!=\!R_\text{lb}({\Dc;\abf}).  \end{align}

\emph{Case 2) $\tilde{N}(\dbf)=2$}:
 In this case, two users request two different files, \ie $\Dc=\{\phi(1),\phi(2)\}$. By the popularity-first placement in \eqref{equ:popu_fir}, we have $a_{\phi(1),l}\ge a_{\phi(2),l}$, $l=1,2$. Following this, for $K=2$ and $\tilde{N}(\dbf)=2$, $R_\text{lb}({\Dc};\abf)$ in \eqref{equ:bnd_5} is given by
 \begin{align}
 R_\text{lb}({\Dc};\abf)=a_{\phi(1),0}+a_{\phi(2),0}+a_{\phi(1),1}.
 \end{align}
For   $R_\text{MCCS}({\dbf};\abf)$ in \eqref{equ:achiev_conv_1}, we have
\begin{align}\label{ach_obj_L00}
&R_{\text{MCCS}}(\dbf;\abf) =\sum_{l=0}^{1}\sum_{i=1}^{2}\sum_{\Sc \in \tilde{\Ac}^{l+1}_i}\bar{a}_{\psi(i),l}^\sSc.
\end{align}
Note that by definition, $\tilde{\Ac}^{2}_2$ contains those user subsets of size two  that include $\psi(2)$ but not $\psi(1)$. However, for $K=2$, when excluding $\psi(1)$, the size of the user subset can only be $1$. Thus, we have $\tilde{\Ac}^{2}_2=\emptyset$ for $K=2$. In addition, we have
$\tilde{\Ac}^{1}_1=\{\psi(1)\}$, $\tilde{\Ac}^{1}_2=\{\psi(2)\}$, and $\tilde{\Ac}^{2}_1=\{\psi(1),\psi(2)\}$. Thus,   $R_{\text{MCCS}}(\dbf;\abf)$ in \eqref{ach_obj_L00} is given by
\begin{align}
R_{\text{MCCS}}(\dbf;\abf) &=\sum_{\Sc \in \tilde{\Ac}^{1}_1}\bar{a}_{\psi(1),0}^\sSc+\sum_{\Sc \in \tilde{\Ac}^{1}_2}\bar{a}_{\psi(2),0}^\sSc+\sum_{\Sc \in \tilde{\Ac}^{2}_1}\bar{a}_{\psi(1),1}^\sSc \nn \\
&=a_{d_{\psi(1)},0}+a_{d_{\psi(2)},0}+a_{d_{\psi(1)},1} \nn\\
 &= a_{\phi(1),0}+a_{\phi(2),0}+a_{\phi(1),1}\nn\\
 &=R_\text{lb}({\Dc};\abf)\label{equ:gap_K2_2}
\end{align}
where the second equality is due to \eqref{a:psi_order}--\eqref{equ:ach_obj_L2}, and the third equality is because of $d_{\psi(i)}={\phi(i)}$.
 From \eqref{equ:gap_K2_11} and \eqref{equ:gap_K2_2}, we conclude that for $K=2$, $R_{\text{MCCS}}(\dbf;\abf)\!=\!R_\text{lb}({\Dc};\abf)$, for $\abf\in\Qc$. Thus, {\bf P0} and {\bf P2} are equivalent.  By Theorem \ref{thm:lbequ}, {\bf P0} and {\bf P1} are equivalent for $K=2$, and we complete the proof.
\endIEEEproof

 For $K=2$, Theorem~\ref{thm_K2} shows that the lower bound given by {\bf P1} is tight. It also indicates two types of optimality for the MCCS: 1) the optimality of the popularity-first cache placement for the MCCS; and 2) the optimality of the MCCS  for caching with uncoded placement. The tight bound enables us to characterize the exact memory-rate tradeoff for caching with uncoded placement. Also, as discussed in Section~\ref{subsec:delivery}, zero-padding is commonly used in constructing  coded messages. The optimality of the MCCS reveals that there is no loss of optimality to use zero-padding in  the MCCS  for the coded message.

\subsection{Region $2$: $K>2$, $\tilde N(\dbf)=K$}\label{sec:peak-load}
When $K\le N$, this region may occur if every user requests  a different file, \ie $|{\Dc}|=|\Uc|=\tilde{N}(\dbf)=K$.
Note that under this condition, the probability of each file being requested has changed.    Let $p_{i|K}$ denote the conditional probability of file $i$ being requested, given $\tilde{N}(\dbf) = K$. Then, in this case,  $\bar{R}_\text{lb}(\abf)$ in \eqref{equ:converse_obj_pf} of {\bf P2} is rewritten as
\begin{align}\label{equ:peak_lb}
\bar{R}_\text{lb}(\abf)&=\sum_{{\Dc}\subseteq\Nc^K} \sum_{\dbf\in\Tc({\Dc})}\prod_{k=1}^{K}p_{d_k|K} R_\text{lb}({\Dc};\abf),
\end{align}
and $\bar{R}_\text{MCCS}(\abf)$ in \eqref{equ:avg_R_MCCS} for the MCCS is rewritten as
\begin{align}\label{equ:peak_obj}
\bar{R}_\text{MCCS}(\abf)&\!=\!\sum_{{\Dc}\subseteq\Nc^K} \sum_{\dbf\in\Tc({\Dc})}\prod_{k=1}^{K}p_{d_k|K}R_{\text{MCCS}}(\dbf;\abf).
\end{align}
Comparing the expressions in \eqref{equ:peak_lb} and \eqref{equ:peak_obj} in  {\bf P0} and {\bf P2}, respectively, we have the following result.

\begin{theorem}\label{thm_peak}
For the caching problem of $N$ files with distribution $\pbf$ and local cache size $M$, in Region 2,  the optimized MCCS in {\bf P0} attains the popularity-first-based lower bound   given by {\bf P2}.
\end{theorem}
\IEEEproof
 To prove the result,  we only need to show that $R_\text{lb}({\Dc;\abf})=R_{\text{MCCS}}(\dbf;\abf)$, for $\abf\in\Qc$ and $\tilde{N}(\dbf) = K$.
Consider $R_{\text{MCCS}}(\dbf;\abf)$ in \eqref{equ:achiev_conv_1}. Since every user requests  a distinct file,  the leader group includes all users and can be written as  $\Uc=\{\psi(1),\ldots,\psi(K)\}$. Thus, for any user subset $\Sc$, we have $\Sc \subseteq \Uc$.  From the definition of $ \tilde{\Ac}^{l+1}_i$ below   \eqref{equ:ach_obj_L3}, this means that for any $\Sc\in \tilde{\Ac}^{l+1}_i$,  $\psi(i)\in \Sc \subseteq \Uc$, and $d_{\psi(i)}$ is the most popular file requested in $\Sc$.  By \eqref{a:psi_order}, we have  $\max_{k\in\Sc }a_{d_k,l}=a_{d_{\psi(i)},l}$. Thus, $\bar{a}_{\psi(i),l}^\sSc$ in \eqref{equ:ach_obj_L3} is given by
\begin{align}
\bar{a}_{\psi(i),l}^\sSc=\max_{k\in\Sc }a_{d_k,l}=a_{d_{\psi(i)},l},   \quad \text{for~} \Sc\in \tilde{\Ac}^{l+1}_i.
\end{align}
Since $|\tilde{\Ac}^{l+1}_{i}|= \binom{K-i}{l}$,  $R_{\text{MCCS}}(\dbf;\abf)$ in \eqref{equ:achiev_conv_1}  is given by
 \begin{align}
 R_{\text{MCCS}}(\dbf;\abf) &=\sum_{l=0}^{K-1}\sum_{i=1}^{K}\binom{K-i}{l}a_{d_{\psi(i)},l}\nn\\
&=\sum_{l=0}^{K-1}\sum_{i=1}^{K}\binom{K-i}{l}a_{\phi(i),l},
 \end{align}
which is the same as $R_\text{lb}({\Dc;\abf})$ in
\eqref{equ:bnd_5} for $\tilde{N}(\dbf)=K$. Thus, we  conclude that $\bar R_\text{lb}(\abf)=\bar R_\text{MCCS}(\abf)$ in Region 2.
 \endIEEEproof

Following Theorem~\ref{thm_peak}, we provide a few conclusions below.

\subsubsection{The optimality of Condition C1}
Note that in Region 2, the MCCS and the CCS are  identical. Recall from Section~\ref{subsec:delivery} that the MCCS uses a modified coded delivery strategy: it removes the redundant coded messages in that of the CCS when there are redundant requests. In Region 2, since all file requests are distinct,  there is no redundant
message in the delivery phase. As a result, the MCCS is the same as the CCS. 

For the CCS, Condition C1 has been proven to be the property of the optimal cache placement for the CCS~\cite{Jin&Cui:Arxiv2017}. Recall that we have imposed Condition C1   in Section~\ref{III.A} to simplify  the cache placement problem for the MCCS.  
The above discussion,  along with the result in Region 1, reveals  that  in Regions 1 and 2,  Condition C1 in fact holds true for the optimal cache placement of the MCCS, \ie the size of each subfile $W_{n,\Sc}$ in file $W_n$ is the same for the user subset of the same size $|\Sc|$. The result is summarized in the following corollary.
\begin{corollary}\label{col1}
Condition C1 is the property of the optimal cache placement  for the MCCS in Regions 1 and 2.
\end{corollary}
Note that the optimality of Condition C1 for the MCCS has only been  demonstrated through numerical results in \cite{Jin&Cui:Arxiv2018}. We show    in Corollary~\ref{col1}   that this property holds for the MCCS in Regions 1 and 2.
\subsubsection{The optimality of popularity-first placement}\label{subsec:PF_opt} Recall from Remark~\ref{remark1} that the popularity-first placement is proven to be optimal for the CCS. Thus, along with the result in Region 1, we have the following corollary.
\begin{corollary}\label{col3}
In Regions 1 and 2, the optimal cache placement for the MCCS  is a popularity-first placement, \ie there is no loss of optimality by restricting $\abf \in \Qc$ in {\bf P0}.
\end{corollary}
\IEEEproof
In Region 1, the claim immediately follows Theorem~\ref{thm_K2}, as discussed at the end of Section~\ref{sec:two-user}. In Region 2, the MCCS is the same as the CCS. It has been shown that the optimal placement for the CCS is a popularity-first placement~\cite{Jin&Cui:Arxiv2017}. Thus, the claim  immediately follows for the MCCS in Region 2.
\endIEEEproof

\subsubsection{Effect of zero-padding in coded caching}\label{remark:zeropadding}
As mentioned in Remark~\ref{remark2}, for  nonuniform file popularity, zero-padding is  a common technique to form the coded messages for both the CCS and the MCCS. However, its impact on the performance has never been discussed or studied. The tight lower bounds shown in Theorems \ref{thm_K2} and  \ref{thm_peak} indicate that the use of zero-padding does not cause any loss, as stated below.
\begin{corollary}\label{col2}
Zero-padding in the coded messages   incurs no loss of optimality for the MCCS in Regions 1 and 2.
\end{corollary}

{Note that  for Case 1  in Region 1 discussed in Section~\ref{sec:two-user}, when both users request the same file, it is clear from \eqref{equ:zero_pad} that zero-padding is not used in the coded message. However, for Case 2 in Region 1, when two users request two different files, zero-padding may be applied in the coded messages. Similarly, in Region 2, for $K$ users requesting $K$ different files,  each coded message contains subfiles of different requested files, and thus, the message may need to be zero-padded.}

\subsubsection{The optimality of the CCS}
The above analysis focuses on the MCCS. {Since the MCCS and the CCS are the same in Region 2, the result in Theorem~\ref{thm_peak}
also leads to several conclusions on the optimality of the  CCS below, which has not been shown in the literature.}
\vspace*{.5em}

\emph{4.i) {Zero-padding}:}\label{pa:zero-padding}
Note that the  delivery strategy of the CCS  does not distinguish whether file requests are the same or different, \ie it treats all the requested files as distinct files to form the coded messages.\footnote{The CCS was originally proposed for the worst-case peak rate consideration, where all file requests are distinct.} This means that  for any  demand vector $\dbf$, the CCS is  effectively equivalent to the case  when $\tilde{N}(\dbf)=K$ for the MCCS in Region 2, where all users request different files.   In other words, the average  rate of the CCS (averaged over \emph{all} $\dbf$'s) is equal to that of the MCCS in Region 2.\footnote{This result should not be confused with the conclusion that the CCS and the MCCS being the same in Region 2 in Section~\ref{subsec:PF_opt}. Here, there  may be multiple requests for the same file, although the CCS does not distinguish them. In Section~\ref{subsec:PF_opt}, the comparison is restricted to Region 2 where all file requests are indeed distinct.}

Assuming all files are \emph{treated} as distinct in the delivery phase, we can also construct a lower bound under popularity-first placement. Given how {\bf P2}  in Region 2 is formulated, this lower bound is  equivalent to {\bf P2}. By Theorem~\ref{thm_peak}, it follows that the average rate of the CCS attains the popularity-first-based  lower bound,  assuming all files are treated as distinct. Since the popularity-first placement is optimal for the CCS under nonuniform file popularity \cite{Jin&Cui:Arxiv2017}, this tight lower bound also means that   zero-padding used in the CCS incurs no loss.
We state this conclusion below.
\begin{corollary}\label{col4}
Using zero-padding for the  CCS incurs no loss of optimality under nonuniform file popularity.
\end{corollary}

\emph{4.ii) The optimality of Condition C1:}   Following the discussion above, the tight lower bound also implies that Condition C1 is the property of the optimal cache placement for the CCS. This is by the similar argument used in the zero-padding discussion. Although this optimality has been proven in \cite{Jin&Cui:Arxiv2017}, the method used there is more involved. Our results  in Theorem~\ref{thm_peak} provides a simpler alternative  proof of this result.
\vspace*{.5em}

\emph{4.iii) The optimality of the CCS:} Following the  discussion in a), since the CCS attains the lower bound given by {\bf P2}, we also conclude that \emph{the CCS is optimal in terms of the average rate for caching under popularity-first placement, if all file requests are distinct (\ie the worst-case)}. Note that in the literature of caching with uncoded placement, for uniform file popularity, the optimality of the CCS   in terms of the worst-case peak rate in the case of $K\le N$ has been proven \cite{Wan2016On1,Wan2016On2}. For nonuniform file popularity, although  many existing works study the cache placement for the CCS   \cite{Niesen&Maddah-Ali:TIT2017,hachem2017coded,Ji&Order:TIT17,Zhang&Coded:TIT18,Daniel&Yu:TIT19,Jin&Cui:Arxiv2017,Saberali&Lampe:TIT20,Deng&Dong:TIT22}, the optimality of the CCS in this case has never been studied or known.
Our result sheds some light on the optimality of the CCS under nonuniform file popularity.

\subsection{Region $3$: $K>2$, $\tilde N(\dbf)<K$}\label{sec:average-load}
This region reflects the  scenario when there exist multiple users request the same file. In the following, we  show that,  in general, there may exist a gap between $R_\text{MCCS}(\dbf;\abf)$ and $R_{\text{lb}}({\Dc};\abf)$,  for $\abf\in\Qc$.
The main cause of the gap is the zero-padding used in the MCCS.

Examining the expressions of  $R_\text{MCCS}(\dbf;\abf)$ in \eqref{equ:achiev_conv_1} and     $R_\text{lb}({\Dc};\abf)$ in \eqref{equ:bnd_5}, we see that they contain the same number of  coded messages, which is  $\sum_{l=0}^{K-1}\sum_{i=1}^{\tilde N(\dbf)}\binom{K-i}{l}$. The only difference between  $R_\text{MCCS}(\dbf;\abf)$ and $R_{\text{lb}}({\Dc};\abf)$ is the size of each coded message $|C_{\Sc}|$, \ie $\bar{a}^\sSc_{\psi(i),l}$ and $a_{\phi(i),l}$.  Thus, we need to examine whether  $\bar{a}^\sSc_{\psi(i),l}$ is the same as $a_{\phi(i),l}$, for any $\Sc$. To better explain our result, in the following, we first use an example to show that   $\bar{a}^\sSc_{\psi(i),l}$ and $a_{\phi(i),l}$ may be different, which is due to zero-padding.

{\bf Example:} Assume that  two users request file $\phi(1)$, the most popular file in the requests. One user is in the leader group $\Uc$, denoted by  $\psi(1)$ and the other from a redundant group, denoted by $k'\notin \Uc$, where $d_{k'}=\phi(1)$. From  $R_\text{lb}({\Dc};\abf)$ in \eqref{equ:bnd_5},  for  all $\binom{K-2}{l}$ user subsets that include user $\psi(2)$ but not user $\psi(1)$, the size of  coded messages corresponding to these subsets is always $a_{\phi(2),l}$ ($d_{\varphi(2)}=\phi(2)$).
Now, for $R_{\text{MCCS}}(\dbf;\abf)$ in \eqref{equ:achiev_conv_1}, consider user subset $\Sc$
that includes users $\psi(2)$ and $k^{\prime}$ but not user $\psi(1)$. From \eqref{equ:ach_obj_L2}, the size of coded messages for $\Sc$  is $\bar{a}^{\Sc}_{\psi(2),l}=a_{d_{k'},l}=a_{\phi(1),l}$, due to zero-padding.
Since $a_{\phi(1),l}\ge a_{\phi(2),l}$, in this case,  zero-padding by the MCCS results in a longer coded message for user subsets that include user $k^{\prime}$ but not the leader user $\psi(1)$, as it always zero-pads the subfile to the longest one in the subset.

Similar to the above example, in general,    $\bar{a}^\sSc_{\psi(i),l}$ and $a_{\phi(i),l}$ may be different  for a coded message corresponding to user subset $\Sc$, where $\Sc$ includes a user from a redundant group who requests a file that is more popular than the rest requested by  all other users in $\Sc$. For the MCCS using the popularity-first placement in \eqref{equ:popu_fir}, the coded message is zero-padded to the size of the subfile requested by that user from the redundant group (the largest). In contrast, for the lower bound $R_{\text{lb}}({\Dc};\abf)$, the size of the coded message is that of the subfile of the most popular file ($\phi(i)$, for some $i$) among files requested by those users   in the leader group, \ie  $\Sc \cap \Uc$. This  mismatch between $R_{\text{MCCS}}(\dbf;\abf)$ and $R_{\text{lb}}({\Dc};\abf)$  leads to a possible gap between the average rate of the optimized MCCS and the lower bound in {\bf P2}.
To further quantify the difference between
$R_\text{MCCS}(\dbf;\abf)$ and $R_{\text{lb}}(\Dc;\abf)$, we re-express $R_\text{MCCS}(\dbf;\abf)$ in the following lemma.
\begin{lemma}\label{lemma_achi}
  For any demand vector $\dbf$, let $\hat N(i)$ denote the total number of redundant requests for files $\{\phi(1),\ldots,\phi(i)\}$, for $i=1,\ldots,\tilde{N}(\dbf)$, and  $\hat N(0)=0$. Then,  $R_{\text{MCCS}}(\dbf;\abf)$ in \eqref{equ:achiev_conv_1} can be rewritten as
\begin{align}
R_{\text{MCCS}}(\dbf;\abf)=&\sum_{l=0}^{K-1}\sum_{i=1}^{\tilde{N}(\dbf)}\left[ \sum_{j=i}^{\tilde N(\dbf)}\binom{K-j-\hat N(i-1)}{l}\right.\nn\\
&\left. \!-\!\sum_{j=i+1}^{\tilde N(\dbf)}\binom{K-j-\hat N(i)}{l} \right] a_{\phi(i),l}.\label{equ:aver_load_4}
\end{align}
\end{lemma}
\IEEEproof
See Appendix \ref{Proof:lemma_archi}.
\endIEEEproof

The expression in  \eqref{equ:aver_load_4}  clearly shows the difference between   $R_{\text{MCCS}}(\dbf;\abf)$ and   $R_{\text{lb}}({\Dc};\abf)$  in \eqref{equ:bnd_5}: the gap between the two is determined by $\hat N(i),i=1,\ldots,\tilde{N}(\dbf)$, \ie the number of redundant  requests for  files in $\Dc$. From this analysis, we identify the following two special cases where $R_{\text{MCCS}}(\dbf;\abf)=R_{\text{lb}}({\Dc};\abf)$:\\[-.75em]

\emph{{Case i):} $|\Dc|=|\Uc|=\tilde{N}(\dbf)=1$}. In this case, all  users request the same file. We have $a_{d_{\psi(1)},l}=\ldots=a_{d_{\psi(K)},l}=a_{\phi(1),l}$, since only one file is requested.
First, we note that $R_{\text{lb}}(\Dc;\abf)$ in  \eqref{R_lblb}  for {\bf P1} and \eqref{equ:bnd_5}  for {\bf P2} are equivalent, given by 
\begin{align}
R_\text{lb}(\Dc;\abf)=\sum_{l=0}^{K-1}\binom{K-1}{l} a_{\phi(1),l}.
\end{align} 
Following this, \eqref{equ:aver_load_4} is given by
\begin{align}
R_{\text{MCCS}}(\dbf;\abf)=&\sum_{l=0}^{K-1}\binom{K-1}{l} a_{\phi(1),l}=R_\text{lb}(\Dc;\abf).\nn
\end{align}

Based on the above, we conclude  that in fact {\bf P0}, {\bf P1} and {\bf P2} are all equivalent in this case. To explain this result, note that as discussed below Corollary~\ref{col2}, when only one file is requested by all the users, zero-padding is not used in the coded message. As a result, zero-padding is avoided, and the optimized MCCS remains to be optimal in this case. 

\emph{{Case ii):} $\hat N(i)=0$, $i=1,\ldots,\tilde N(\dbf)-1$}. In this case, only file $\phi(\tilde{N}(\dbf))$,  \ie the least popular file in $\Dc$, has redundant requests. In other words, all the users in the redundant group request file $\phi(\tilde{N}(\dbf))$. From  \eqref{equ:aver_load_4} and \eqref{equ:bnd_5}, for {$\abf\in\Qc$}, it is straightforward to show that
\begin{align}
R_{\text{MCCS}}(\dbf;\abf)=\sum_{l=0}^{K-1}\sum_{i=1}^{\tilde{N}(\dbf)}\binom{K-i}{l}a_{\phi(i),l}=R_\text{lb}(\Dc;\abf).\nn
\end{align}
{Thus,  {\bf P0} and {\bf P2} are equivalent in this case. To explain this, we see that in this case,  the redundant file requests are ``controlled,"  \ie the redundant requests are for the least popular file $\phi(\tilde{N}(\dbf))$ in $\Dc$. By the popularity-first placement condition in \eqref{equ:popu_fir}, we have $a_{1,l}\ge a_{2,l}\ge\ldots\ge a_{\phi(\tilde{N}(\dbf)),l}$ for $l=1,\ldots,K$. As a result, the size of the coded message $|C_{\Sc}|$, for any user subset $\Sc$, is always the size of the subfile of the most popular file requested by the  users in $\Sc$ that belong to the leader group $\Uc$, \ie  $\Sc\cap \Uc$. Thus, ~\eqref{equ:ach_obj_L3} is given by\begin{align}
\bar{a}_{\psi(i),l}^\sSc=\max_{k\in\Sc }a_{d_k,l}=a_{d_{\psi(i)},l}=a_{{\phi(i)},l},   \ \text{for~} \Sc\in \tilde{\Ac}^{l+1}_i.
\end{align} 
We see from the above that although zero-padding is used, the size of the coded message $\bar{a}^\sSc_{\psi(i),l}$ is the same as $a_{\phi(i),l}$ for any $\Sc$. As a result, there is no  loss of optimality caused by zero-padding in this case.}

Finally, we point out that although $\bar R_{\text{MCCS}}(\abf)$ and $\bar R_{\text{lb}}(\abf)$ are generally not the same due to the existence of  redundant requests as discussed above, the optimal placement solution $\abf^*$ to {\bf P0}, {\bf P1}, and {\bf P2} can still be the same for some values of   $M$ and $\pbf$. This also leads to the following case where  $\bar{R}_{\text{MCCS}}(\abf)=\bar{R}_{\text{lb}}(\abf)$:\\[-.75em]

\emph{{Case iii):} If {\bf P0}, {\bf P1} and {\bf P2} result in the same optimal solution $\abf^*$ that  satisfies $\abf^*_1=\cdots=\abf^*_N$, then $\bar R_{\text{MCCS}}(\abf^*)=\bar R_{\text{lb}}(\abf^*)$.}
 In this case, since $a_{1,l}^*=\ldots=a_{N,l}^*$, for any $l$, all the subfiles in a coded message $C_{\Sc}$ in \eqref{equ:popu_fir} for user subset $\Sc$ have the same size. Thus, there is no zero-padding and  no potential waste. One obvious example is the special case of uniform file popularity, where the optimal $\abf^*_n$'s  are  all identical and  the same for {\bf P0}, {\bf P1}, and {\bf P2}.  In this case, the same result $\bar{R}_{\text{MCCS}}(\abf^*)=\bar{R}_{\text{lb}}(\abf^*)$ has been shown in~\cite{Yu&Maddah-Ali:TIT2018}. Below, we provide a simple proof of our statement.

Since  $a^*_{1,l}=\ldots=a^*_{N,l}$, $l=0,\ldots,K$,  $\bar{a}_{\psi(i),l}^\sSc$ in \eqref{equ:ach_obj_L3} can be written as
\begin{align}
\bar{a}_{\psi(i),l}^\sSc=\max_{k\in\Sc }a^*_{d_k,l}=a^*_{1,l},   \quad \text{for~} \Sc\in \tilde{\Ac}^{l+1}_i.
\end{align}
Since $|\tilde{\Ac}^{l+1}_{i}|= \binom{K-i}{l}$, $R_{\text{MCCS}}(\dbf;\abf^*)$ in \eqref{equ:achiev_conv_1} is written as
\begin{align}
 R_{\text{MCCS}}(\dbf;\abf^*) &=\sum_{l=0}^{K-1}\sum_{i=1}^{\tilde N(\dbf)}\binom{K-i}{l}a^*_{1,l}.
 \end{align}
Similarly,  $R_\text{lb}(\Dc;\abf)$ in \eqref{R_lblb} of {\bf P1} and \eqref{equ:bnd_5} of {\bf P2}  can both  be rewritten as
\begin{align}
R_\text{lb}(\Dc;\abf^*)= \sum_{l=0}^{K-1}\sum_{i=1}^{\tilde{N}(\dbf)}\binom{K-i}{l}a^*_{1,l}=R_{\text{MCCS}}(\dbf;\abf^*).
\end{align}
Thus, from \eqref{equ:P1_obj} and \eqref{equ:avg_R_MCCS}, we conclude that
for $\abf^*_1=\ldots=\abf^*_N$,
\begin{align}\label{equ:R_symmetric}
\hspace*{-1em}\bar R_{\text{MCCS}}(\abf^*)&=\bar R_{\text{lb}}(\abf^*)\nn\\
  &=\sum_{\Dc\subseteq\Nc}\!\sum_{\dbf\in \Tc({\Dc})}\!\prod_{k=1}^{K}p_{d_k}\!\!\sum_{l=0}^{K-1}\sum_{i=1}^{\tilde{N}(\dbf)}\binom{K-i}{l}a^*_{1,l}.
\end{align}

\begin{Remark}
In Region 3, as discussed earlier, when there are redundant file requests, zero-padding  the message to the largest subfile may lead to a loss in the average rate, {and the optimality of the MCCS is uncertain. Despite this,  we have identified three special cases where the optimized MCCS  still achieves the lower bounds. In these cases, either  zero-padding is avoided, or the redundant file requests are carefully controlled, such that the potential loss by zero-padding is eliminated. These findings may further guide us in designing caching schemes to reduce the loss  caused by zero-padding.} One possible solution is to create as many subfiles of equal size as possible during the placement phase. Coincidentally, such an approach was also exploited in~\cite{Sahraei2019TheOptimal} for $N=2$  files, where a placement scheme was proposed to  create equal subfile size and was shown to be an optimal caching scheme with uncoded placement for two files.\end{Remark}

\section{Optimal Cache Placement for the MCCS}\label{sec:MCCS_placement}
 Due to the more complicated delivery scheme by the MCCS, finding the optimal cache placement for the MCCS under nonuniform file popularity is  challenging, and the problem has not been solved.
The optimal cache placement for the CCS under nonuniform file popularity has recently been obtained in~\cite{Deng&Dong:TIT22}. In this section, through reformulating {\bf P0}, we show that the  cache placement problem  has a similar structure to that for the CCS.  As a result,  the optimal cache placement structure for the MCCS inherits that for the CCS  characterized in~\cite{Deng&Dong:TIT22}. Extending the results from the CCS, we  present a simple algorithm to compute the optimal cache placement solution for the MCCS.
In the following, we first reformulate {\bf P0} and then describe the optimal cache placement solution structure.

\subsection{Reformulation of {\bf P0}}\label{sec:MCCS_reformulation}
It is straightforward to see that at the optimum, the cache memory is  fully utilized, and the local cache constraint \eqref{Constraint2} is attained with equality, which can be replaced by
  \begin{align}\label{Constraint1.1_eq}
  \sum_{n=1}^{N}\sum_{l=1}^{K}{K-1 \choose l-1}a_{n,l} = M.
  \end{align}
        
Next, for any popularity-first placement $\abf\in\Qc$,  constraint \eqref{Constraint_gt0} can be equivalently replaced by the following two constraints~\cite[Lemma 1]{Deng&Dong:TIT22}:
\begin{align}
a_{1,0}\geq 0\quad \text{and}\quad a_{N,l}\geq 0,\quad l\in\Kc. \label{Constraint_a_Nl}
\end{align}

Finally, for $\abf\in\Qc$,  the expression of the average rate $\bar{R}_\text{MCCS}(\abf)$ in \eqref{avgR_MCCS} can  be  simplified as \cite{Jin&Cui:Arxiv2018}\footnote{
The expression in \eqref{avgR_MCCS}  can be simplified to   \eqref{equ:AverageRate} by utilizing the properties of the popularity-first placement with ordered $a_{n,l}$'s  to eliminate the max operation in \eqref{avgR_MCCS}. The expression of $\bar R_\text{MCCS}(\abf)$ in   \eqref{equ:AverageRate} can be evaluated in polynomial time instead of the exponential time required in  \eqref{avgR_MCCS}, which simplifies {\bf P0}. }
\begin{align}\label{equ:AverageRate}
    &\!\!\!\!\bar{R}_\text{MCCS}(\abf)=\nn\\
    &\sum_{l=0}^{K-1}\!\!\binom{K}{l\!+\!1}\!\sum_{n=1}^{N}\left(\!\! \left(\sum_{n^{\prime}=n}^{N}p_{n^{\prime}} \!\!\right)^{l+1}\! \!\!\!\!-\! \left(\sum_{n^{\prime}=n+1}^{N}p_{n^{\prime}} \right)^{l+1} \right)\!a_{n,l}\nn\\
    &-\!\!\!\sum_{u=1}^{\min\{N,K\}}\!\sum_{l=0}^{K-\!u-\!1}\!\!\!\binom{K-\!u}{l+\!1}\!\sum_{i=1}^{K-u}\!\binom{K\!-u\!-i}{l}\!\sum_{n=1}^{N}P_{i,u,n} a_{n,l}
\end{align}
where $P_{i,u,n}$ is  the joint probability of i) having  $u$ distinct file requests; and ii) file $W_n$ being the $i$-th least popular file among files requested by all the users that are not in the leader group $\{d_k: k \in \Kc\backslash\Uc\}$. The expression of
$P_{i,u,n}$ is derived in \cite{Jin&Cui:Arxiv2018}, which is lengthy and non-essential in developing our results. Therefore, we omit it here but only point out that $P_{i,u,n}$  is not a function of $\abf$.

The expression of    $\bar{R}_\text{MCCS}(\abf)$   in   \eqref{equ:AverageRate} is a weighted sum of $a_{n,l}$'s.
Define $\gbf_n\triangleq [g_{n,0},\ldots,g_{n,K}]^{\mathrm{T}}$, with\begin{align}\label{g_nl}
&g_{n,l}\triangleq\binom{K}{l+1}\left( \left(\sum_{n^{\prime}=n}^{N}p_{n^{\prime}} \right)^{l+1}-\left(\sum_{n^{\prime}=n+1}^{N}p_{n^{\prime}} \right)^{l+1}\right)\nn\\
&-\sum_{u=1}^{\min\{N,K\}}\binom{K-u}{l+1}\!\sum_{i=1}^{K-u}\!\binom{K-u-i}{l}\!P_{i,u,n}.
\end{align}
Also, from \eqref{Constraint1.1} and \eqref{Constraint1.1_eq}, define $\bbf\triangleq[b_0,\dots,b_K]^T$ with $b_l \triangleq{K \choose l}$, and  $\cbf\triangleq[c_0,\dots,c_K]^T$ with $c_l\triangleq{K-1 \choose l-1}$. The cache placement  optimization problem {\bf P0}    can be reformulated into the following equivalent LP problem
\begin{align}
\textrm{\bf P3:}\; \min_{\abf\in\Qc}  \quad
                  & \sum_{n=1}^{N} \gbf_n^\mathrm{T} \abf_n\ \quad\nn\\
\textrm{s.t.} \quad
                   &\eqref{Constraint_a_Nl},\ \text{and} \nn\\
                   &\bbf^\mathrm{T} \abf_n=1, \;                        n\in\Nc, \label{Constraint_SumTo1}\\
                   &\sum_{n=1}^{N}\cbf^\mathrm{T} \abf_n= M\label{Constraint_SumLeM}.
\end{align}

\subsubsection{Connection to the cache placement  problem for  the CCS} As described in Section~\ref{subsec:delivery}, the MCCS only delivers  the non-redundant messages, while the CCS delivers the coded messages corresponding to all the user subsets.
The expression of $\bar{R}_\text{MCCS}(\abf)$ in \eqref{equ:AverageRate} explicitly shows this difference by grouping all the messages into the first term and  the redundant messages into the second term.
The cache placement optimization problem for the CCS was formulated in~\cite[problem {\bf P2}]{Deng&Dong:TIT22}. It  essentially  minimizes the first term in \eqref{equ:AverageRate}, with all the constraints on $\abf$ being the same as in  {\bf P3}, except that the expression of $\gbf_n$ is different. Based on this structural similarity, the structural properties of  the placement obtained for the CCS can be straightforwardly extended to {\bf P3} for the MCCS.

Since the cache placement result is the direct extension from that of the CCS in \cite{Deng&Dong:TIT22}, in the following, we focus on describing the placement structure and omit the details of derivations or proofs.

\subsection{Optimal Cache Placement: Solution\ Structure}\label{sec:MCCS_structure}
A major challenge in solving {\bf P3} is that the placement vectors in the optimal cache placement $\abf$ can be all different, depending on the file popularity distribution. It turns out that the number of distinct placement vectors in the optimal $\abf$ is limited to at most three.
We first define \emph{file group} below.

\begin{myDef}
[File group] A file group is a subset of $\Nc$ that contains all files with the same cache placement vector, \ie for any two files $W_n$ and $W_{n^{\prime}}$, if  their placement vectors  $\abf_{n}=\abf_{n^{\prime}}$, then they belong to the same file group.
\end{myDef}

The first structural property, in terms of file groups, of the optimal cache placement for the MCCS is provided in the following theorem.

\begin{theorem}\label{The3Groups}
For $N$ files with any file popularity distribution  $\pbf$, and for any $K$ and $M\le N$,  there are at most three file groups under the optimal cache placement  $\{\abf_n\}$ for  {\bf P3}.
\end{theorem}

\IEEEproof
The only differences between {\bf P3} and the the cache placement optimization problem for the CCS  in~\cite{Deng&Dong:TIT22} are the expressions of $\gbf_n$, which does not affect the arguments used in the proof of \cite[Theorem 1]{Deng&Dong:TIT22}. Thus,  the same result  straightforwardly  applies to the optimal solution of {\bf P3} for the MCCS.
\endIEEEproof

Theorem~\ref{The3Groups} indicates that, regardless of  $N$, $\pbf$, and $M$, there are at most three unique values among the optimal  placement  vectors $\{\abf_n\}$.  This leads to three possible cases:  one, two, or three file groups. In the following, we provide the optimal  placement  solution $\{\abf_n\}$ for each case.

\subsubsection{One file group}\label{sec:oneGroup}
In this case, the optimal cache placement vectors are identical for all files, \ie $\abf_1=\cdots=\abf_N$.
Under this structure, the cache placement problem is reduced to that under  uniform file popularity (\ie all files have the same cache placement vector), of which the optimal solution has been obtained in closed-form \cite{Deng&Dong:SPAWC19}, which is identical to that for the CCS  under uniform file popularity \cite{Daniel&Yu:TIT19}. Specifically, the optimal placement $\abf_n$, for any file $n\in\Nc$, has at most two nonzero elements, which is given as follows:
\begin{align}\label{one_group:a_opt}
\begin{cases}
a_{n,l_o}=\frac{1+\left\lfloor v\right\rfloor- v}{{K \choose \left\lfloor v\right\rfloor}}, \;
 a_{n,l_o+1}=\frac{v-\left\lfloor v\right\rfloor}{{K \choose \left\lceil v\right\rceil}}, & l_o=\left\lfloor v\right\rfloor  \\
 a_{n,l}=0, & \hspace*{-1.2em} \forall \; l \neq l_o \ \text{or} \ l_o+1
\end{cases}
\end{align}
where  $v\triangleq \frac{MK}{N}$. In particular, when $v$ is an integer, the optimal  $\abf_n$ has only one nonzero element: $a_{l_o}=1/{K \choose l_o}$ for  $l_o =MK/N $, and $a_l=0$,  $\forall \ l\neq l_o$.

\subsubsection{Two file groups}
In this case, there are only two unique placement vectors in $\{\abf_n\}$, \ie
$\abf_1=\cdots=\abf_{n_o} \neq \abf_{n_o+1}=\cdots=\abf_{N}$, for some $n_o\in \{1,\ldots, N-1\}$.
We use $\abf_{n_o}$ and $\abf_{n_o+1}$ to represent the two  unique placement vectors for the first and the second file group, respectively.
Define $\bar{\abf}_{n}\triangleq[a_{n,1},\ldots,a_{n,K}]^T$as  the sub-placement vector in $\abf_{n}$. It specifies  the subfiles stored in the local cache, and $a_{n,0}$ specifies the subfile kept at the server. Let $\bar\abf_n\succcurlyeq_1 {\bf 0}$ denote that there is only one positive element in $\bar\abf_n$, and the rest are all $0$'s.

There are several structural properties of the optimal placement in the two-file-group case. They are all direct extensions from the optimal cache placement of the CCS~\cite{Deng&Dong:TIT22}. We summarize them below.

\begin{proposition}\label{pro:twogroups}
If there are two file groups under the optimal cache placement $\{\abf_n\}$, the following three properties hold:\\
\emph{Property 1~\cite[Proposition 1]{Deng&Dong:TIT22}:} The optimal sub-placement vector  $\bar\abf_{n_o+1}$ for the second file group has at most one nonzero element.\\
\emph{Property 2~\cite[Proposition 2]{Deng&Dong:TIT22}:} If $\bar\abf_{n_o+1}\succcurlyeq_1\mathbf{0}$, then  $\bar\abf_{n_o}$ and $\bar\abf_{n_o+1}$ are different by only one element.\\
\emph{Property 3~\cite[Proposition 3]{Deng&Dong:TIT22}:} If $\bar\abf_{n_o+1}\succcurlyeq_1\mathbf{0}$, then $a_{n_o,0}=0$.
\end{proposition}

Following the properties in Proposition \ref{pro:twogroups}, the optimal placement solution for {\bf P3} can be one of the following two  structures: 1)  $\bar\abf_{n_o+1}=\mathbf{0}$; 2)   $\bar\abf_{n_o+1}\succcurlyeq_1 \mathbf{0}$. For the completeness, we  briefly present the  optimal placement solutions below, referring the derivation details to~\cite{Deng&Dong:TIT22}.

\textbf{Case 1: $\bar\abf_{n_o+1}=\mathbf{0}$}.\label{sec:nonZeroSecond}
This condition means that no cache is allocated to the files in the second group. By \eqref{Constraint1.1}, we have $\abf_{n_o+1}=[1,0,0,\ldots]^T$.
To determine $\abf_{n_o}$ for the first group, we treat the first $n_o$ files as a new database. Then, the cache placement problem is reduced to the one in the one-file-group case in Section~\ref{sec:oneGroup}. Therefore, the solution is the same as in   \eqref{one_group:a_opt}, except that  $N$ is replaced by $n_o$, and thus, $v = MK/n_o$.

Note that this two-file-group case and  the one-file-group case in Section~\ref{sec:oneGroup} can be combined as follows: The optimal $\abf_{n_o}$ is given by \eqref{one_group:a_opt}, for $v = MK/n_o$ with $n_o\in \Nc$.
 The optimal $n_o^*$ depends on $(N,\pbf, M,K)$, which can be determined via a 1-D search over $n_o$ that  gives the minimum $\bar{R}_\text{MCCS}(\abf)$, where $\bar{R}_\text{MCCS}(\abf)$ is computed using the closed-form expression in \eqref{equ:AverageRate}. For the overall algorithm, please refer to \cite[Algorithm~1]{Deng&Dong:TIT22}, with the only exception that the average rate  is computed using  $\bar{R}_\text{MCCS}(\abf)$ in \eqref{equ:AverageRate}.

\textbf{Case 2: $\bar\abf_{n_o+1}\succcurlyeq_1 \mathbf{0}$}.\label{sec:oneZeroSecond}
In this case,  only one  element in $\bar{\abf}_{n_o+1}$ is nonzero. Assume  $a_{n_o+1,l_o}>0$, for some $l_o \in \Kc$, and $a_{n_o+1,l}=0$, $\forall l\neq l_o$, $l\in \Kc$.
By Property 2 in Proposition \ref{pro:twogroups},
the  element that is different between $\bar\abf_{n_o}$ and $\bar\abf_{n_o+1}$ can be either at index  $l_o$ or some $l_1$, for $l_1\neq l_o$. Thus, for $\abf\in\Qc$,  there are only  two possible cases for $(\abf_{n_o},\abf_{n_o+1})$: \emph{2.i)} $a_{n_o,l_o}>a_{n_o+1,l_o}>0$; or
\emph{2.ii)} $a_{n_o,l_1}>a_{n_o+1,l_1}=0$, for some   $l_1\neq l_o$, $l_1\in\Kc$.
We present the solution  in  each of these two  cases:

\emph{Case 2.i)}  $a_{n_o,l_o}>a_{n_o+1,l_o}>0$: In this case, the only different element between  $\bar\abf_{n_o}$ and $\bar\abf_{n_o+1}$ is at position $l_o$, \ie the nonzero element $a_{n_o+1,l_o}$ in $\bar\abf_{n_o+1}$. Also, by Property 3 in Proposition \ref{pro:twogroups},  $a_{n_o,l_o}$ is the only nonzero element in $\abf_{n_o}$, and $a_{n_o,l}=a_{n_o+1,l}=0$, for $\forall l\neq l_o, l\in\Kc$. For $\bar\abf_{n_o+1}$, there are two unknown nonzero elements $a_{n_o+1,0}$ and $a_{n_o+1,l_o}$. Solving the unknown elements in $\abf_{n_o}$ and $\abf_{n_o+1}$ using constraints \eqref{Constraint_SumTo1} and  \eqref{Constraint_SumLeM}, the optimal $(\abf_{n_o},\abf_{n_o+1})$   is given by
 \begin{align}
&\hspace*{-0.5em}\begin{cases}a_{n_o,l_o}=\frac{1}{{K \choose l_o}}, \quad a_{n_o,l}=0, \ \forall \ l\neq l_o \\ 
\displaystyle a_{n_o+1,0}=1-\!\!\left(\!\frac{\frac{KM}{l_o}-n_o}{N-n_o}\!\right), \
a_{n_o+1,l_o}= \frac{1}{{K \choose l_o}}\!\!\left(\!\frac{\frac{KM}{l_o}-n_o}{N-n_o}\!\!\right)
   \\
a_{n_o+1,l}= 0,  \ \forall \ l\neq 0 \ \text{or} \ l_o
\end{cases}  \label{equ:Twogroups2i_2}
\end{align}
where $n_o \in \{1,\ldots, N-1\}$, and the nonzero element position $l_o$ is limited to
$\left\lfloor\frac{KM}{N}\right\rfloor+1 \le  l_o \le \min \left\{K, \left\lceil\frac{KM}{n_o}\right\rceil-1\right\}$.

\emph{Case 2.ii)} $a_{n_o,l_1}>a_{n_o+1,l_1}=0$,  $l_1\neq l_o$: In this case, the only different element between  $\bar\abf_{n_o}$ and $\bar\abf_{n_o+1}$ is at position $l_1$, which is one of the zero elements in $\bar\abf_{n_o+1}$. It follows that $a_{n_o,l_o}=a_{n_o+1,l_o}>0$.
Since $a_{n_o,0}=0$ (by Property 3 in Proposition \ref{pro:twogroups}),
   $\abf_{n_o}$ has two nonzero elements, $a_{n_o,l_o}$ and $a_{n_o,l_1}$, and the rest are all $0$'s.  For $\abf_{n_o+1}$, there are two unknown nonzero elements, $a_{n_o+1,0}$ and $a_{n_o+1,l_o}=a_{n_o,l_o,}$ and the rest are all $0$'s.
Thus, we have three unknown elements $a_{n_o,l_o}=a_{n_o+1,l_o}$, $a_{n_o,l_1}$, and $a_{n_o+1,0}$ to determine in $\abf_{n_o}$ and $\abf_{n_o+1}$. Solving these unknown elements using constraints \eqref{Constraint_SumTo1} and  \eqref{Constraint_SumLeM}, the optimal $(\abf_{n_o},\abf_{n_o+1})$   is given by
\begin{align}
&\hspace*{-1em}\begin{cases}\displaystyle a_{n_o,l_o}=\frac{1}{{K \choose l_o}}\frac{\frac{KM}{l_1}-n_o}{\frac{l_o}{l_1}N-n_o},
\
a_{n_o,l_1}=\frac{1}{{K \choose l_1}}\frac{\frac{l_o}{l_1}N-\frac{KM}{l_1}}{\frac{l_o}{l_1}N-n_o}
\\
a_{n_o,l}=0, \ \forall \ l\neq l_o \ \text{or} \ l_1 \\
\displaystyle a_{n_o+1,l_o}=a_{n_o,l_o}, \ \  \quad  a_{n_o+1,0}= 1-\frac{\frac{KM}{l_1}-n_o}{\frac{l_o}{l_1}N-n_o}  \\
a_{n_o+1,l}= 0, \forall \ l\neq 0 \ \text{or} \ l_o
\end{cases}\label{equ:Twogroup2ii_2}
\end{align}
where positions $l_o$ and $l_1$  satisfy either of the following two conditions: i) $l_o > KM/N$ and $ l_1 < KM/n_o$, or ii) $ l_o<KM/N$ and $l_1 > KM/n_o$.
Note that since $l_o,l_1\le K$,  if $n_o\le M$, only  the condition in i) is valid.

In summary, we can consider Case 2.i) as the special case when $l_1=l_o$. Then,  for given $(n_o,l_o,l_1)$, the closed-form solution  in \eqref{equ:Twogroups2i_2} of Case 2.i), or \eqref{equ:Twogroup2ii_2} of Case 2.ii)
completely determines the optimal $\abf_{n_o}$ and $\abf_{n_o+1}$.
We can  search over all possible values of $n_o\in\{1,\ldots,N-1\}$ and $l_o,l_1\in \Kc$ for the optimal tuple $(n_o,l_o,l_1)$  that gives minimum  $\bar{R}_\text{MCCS}$.  For the overall algorithm, please refer to \cite[Algorithm~2]{Deng&Dong:TIT22}.
\begin{Remark}
We have shown two different structures of the optimal placement in Cases 1 and 2  for the two file groups. In Case 1, all the cache is allocated to the first group. For each file  in this group, all its subfiles are  cached at different users, and the cache placement is identical for these files, regardless of their popularity. In the existing works, the two-file-group strategies proposed in \cite{Ji&Order:TIT17,Zhang&Coded:TIT18} coincide with this structure, although they are proposed for  a decentralized cache placement, and the methods to determine $n_o$ are heuristic or suboptimal. In Case 2, each file in the second group is partially cached at the users and partially remains at the server. This placement structure has never been proposed in the literature. For this placement structure, coding opportunity between the two file groups is explored to minimize the average rate.  
\end{Remark}

\subsubsection{Three file groups}\label{sec:threeGroups}

Under this structure, there are three unique placement vectors in  $\{\abf_n\}$, \ie $\abf_{1}=\ldots=\abf_{n_o}\neq\abf_{n_o+1}=\ldots=\abf_{n_1}\neq\abf_{n_1+1}=\ldots=\abf_{N}$, for $1\le n_o<n_1\le N-1$. We use $\abf_{n_o}$, $\abf_{n_o+1}$ and $\abf_{n_1+1}$ to represent the three unique cache placement vectors for the first, second, and the third file group, respectively.
Following the proof of~\cite[Proposition 4]{Deng&Dong:TIT22}, it is straightforward to show the same result in the following holds for the MCCS as well: All the memory is allocated to the first two file groups and the optimal cache placement vector for the third group is $\abf_{n_1+1}=[1,0,0,\ldots]^T$.

Following the above, we treat those  $n_1$ files in the first two groups as a new database.  The cache placement problem for  these first two groups  is essentially reduced to the previous two-file-group case.
Since $\abf_{n_o+1}\neq\abf_{n_1+1}$, we have $\bar{\abf}_{n_o+1}\neq\bar{\abf}_{n_1+1}=\mathbf{0}$. This means that  $\bar{\abf}_{n_o+1}$ includes at least one nonzero element. Therefore, the cache placement  $(\abf_{n_o}, \abf_{n_o+1})$  belongs to the case of  two file groups  with $\bar{\abf}_{n_o+1}\succcurlyeq_1 \mathbf{0}$ (for the second group) in Case 2 of Section~\ref{sec:oneZeroSecond}. For given $n_1$, the  optimal solution for  $(\abf_{n_o}, \abf_{n_o+1})$ can be obtained from  \eqref{equ:Twogroups2i_2} or \eqref{equ:Twogroup2ii_2}, except that $N$ is replaced by $n_1\in\{2,\ldots,N\}$.

The final optimal $\{\abf_n\}$ is obtained by searching over all possible values of $n_1\in\{2,\ldots,N\}$,\footnote{Further analysis shows that we can limit the range of search for $n_1$ within $n_1 \in \{M+1,\ldots,N-1\}$~\cite{Deng&Dong:TIT22}.} $n_o\in\{1,\ldots,n_1-1\}$,  and $l_o,l_1\in \Kc$ for the optimal tuple $(n_o,n_{1,}l_o,l_1)$  that results in minimum  $\bar{R}_\text{MCCS}$. For the overall algorithm, please refer to \cite[Algorithm~3]{Deng&Dong:TIT22}.
The algorithm simply computes $\bar{R}_\text{MCCS}$ using a closed-form expression for at most $(N-1)(N-2)K^2/2$ times with different values of $(n_o,n_1,l_o,l_1)$, which can be computed efficiently in parallel.
\begin{Remark}
Note that the three-file-group structure for cache placement   described above  has never been proposed in the literature for coded caching. Although \cite{Zhang&Coded:TIT18} has considered  three file groups as a choice in a mixed caching scheme, this three-file-group cache placement  is only  used with uncoded delivery and is only invoked in some rare cases, which is different from coded caching. In contrast, for the  three-file-group structure of the optimal cache placement described above, coding opportunity among three file groups is explored to minimize the average rate. 
\end{Remark}

\subsubsection{The Optimal Cache Placement Solution}\label{subsubsec:opt_placement}
In summary, {by exploring  optimization techniques and  the structural properties of the solution, we can obtain the optimal cache placement solution for the MCCS in  {\bf P0}.} By Theorem~\ref{The3Groups},
 the optimal cache placement problem {\bf P3} is reduced to a search among three possible file grouping structures (from one to three file groups). From Sections~\ref{sec:oneGroup} to \ref{sec:threeGroups}, using the closed-form expressions for $\{\abf_n\}$ and $\bar{R}_\text{MCCS}(\abf)$, we obtain the candidate optimal placement  for each file-group case. The final optimal placement is  the one in these  three solutions that results in the minimum  $\bar{R}_\text{MCCS}(\abf)$. Since the overall algorithm only involves parallel evaluations of closed-form expressions, the algorithm is very simple  with low computational complexity.

Beyond the analytical solution, our results also bring important insight into cache placement under nonuniform file popularity. We have fully characterized the inherent file grouping structure of the optimal cache placement solution for the MCCS. Theorem~\ref{The3Groups} shows that regardless of the file popularity distribution, there are at most three file groups in the optimal cache placement, with files in the same group having the same cache placement. We categorize these three groups as ``most popular," ``moderately popular," and ``non-popular" files. The result indicates that despite different file popularities, the caching method only distinguishes files in these three categories to determine the cache placement strategies. In other words, these three categories reflect the caching strategies: for the ``most popular" file group, cache all their subfiles; for  the  ``moderately popular" file group,  each file is cached only partially, and the rest is stored at the server; and for ``non-popular" file group, the files are not cached but solely stored at the server. How many file groups and for which category, \ie the mapping of files to the three categories, depend on the file popularity distribution $\pbf$ and the ratio of total cache size among users to the server database size $KM/N$. In our numerical results, we will demonstrate these  file group structures in cache placement in Table~\ref{tbl:file_grouping} as the cache size varies. This can be determined by the optimal cache placement solution via our algorithm above.

Finally, we note that the optimal cache placement solution for the MCCS obtained in this section allows us to quantitatively evaluate the gap between the optimized MCCS and the lower bounds in {\bf P1} and {\bf P2} in Region 3 discussed in Section~\ref{sec:average-load}.

\begin{Remark}
We point out that,  for nonuniform file popularity, although the MCCS and the CCS have the same set of  candidate optimal  cache placement structures and solutions, the final optimal cache placement for the two schemes may be different due to different expressions of the average rate (\eg $\bar{R}_\text{MCCS}(\abf)$ in {\bf P0}). We will demonstrate this in our numerical studies in Section \ref{sec:num}. Note that this is in contrast to  uniform file popularity, where the optimal cache placements of the MCCS~\cite{Yu&Maddah-Ali:TIT2018} and the CCS~\cite{Daniel&Yu:TIT19} are exactly the same.
\end{Remark}

\section{Memory-rate Tradeoff For Nonuniform File Popularity and Size}\label{sec:non_length_popu}
In the previous sections, we have focused on files of equal size.\footnote{In practice, files with nonuniform sizes could also be tailored into files with uniform size which are treated separately with different popularities~\cite{pedarsani2016online,Niesen15CodedDelay,Park16Joint}. } In this section, we extend our study on the memory-rate tradeoff in  caching to the more general case where both file popularity and size are nonuniform among files. We extend the cache placement optimization formulation in Section~\ref{sec:prob} to this case and propose a lower bound for caching with uncoded placement. By comparing the average rates of the optimized MCCS and the lower bound, we characterize the exact memory-rate tradeoff for $K=2$ users.

\subsection{The Optimized MCCS }\label{sec:MCCS_length}
Consider each file of different size. We assume that file $W_n$ has  $F_n$ bits. Recall in Section \ref{sec:model} that, for  uniform file size, subfile size $a_{n,l}$ and cache size $M$ are normalized by the file size and defined in the unit of file. For files with different sizes, we remove this normalization. Instead, for each file $n\in\Nc$, we define the size of each subfile in bits: $a_{n,l}\triangleq |W_{n,\Sc}|$, for $|\Sc|=l$.  Likewise, the cache size $M$ is   now defined in bits. Accordingly, we rewrite file partition constraint  \eqref{Constraint1.1} as
\begin{align}\label{non_length_Constraint2}
\sum_{l=0}^{K}{K \choose l}a_{n,l}=F_n, \ n \in \Nc.
\end{align}

With the above redefinitions of $a_{n,l}$ and $M$, the expressions of the cache size constraint \eqref{Constraint2} and the average delivery rate $\bar{R}_{\text{MCCS}}(\abf)$ in \eqref{avgR_MCCS} still remain the same under the nonuniform file popularity and size.
The cache placement  optimization problem for the MCCS  to minimize $\bar{R}_{\text{MCCS}}(\abf)$ under nonuniform file popularity and size is formulated as follows:
\begin{align}
\textrm{\bf P4}: \;\min_{\abf}&\;\;  \bar{R}_{\text{MCCS}}(\abf) \nn\\
 \textrm{s.t.} &\;\;
\eqref{Constraint2},\eqref{Constraint_gt0},  \text{and~} \eqref{non_length_Constraint2}\nn
\end{align}
where $\bar{R}_{\text{MCCS}}(\abf)$ is given in \eqref{avgR_MCCS}, and   the objective and constraint functions are now all expressed in bits.

Note that different from {\bf P0}, which is restricted to the popularity-first placement $\abf\in \Qc$, the problem size of {\bf P4} grows exponentially with $K$. We will not further study the simplification of {\bf P4} and its performance in this paper. For the CCS, a similar problem under nonuniform file popularity and size has been studied, and tractable techniques have been developed  to simplify the optimization problem   with good performances~\cite{Daniel&Yu:TIT19}. The techniques can be adopted here for {\bf P4} for the MCCS, due to the similarity between the two caching schemes.
In the following, we focus on the characterization of the memory-rate tradeoff under nonuniform file popularity and size, which is unknown in the literature.

\subsection{Memory-Rate Tradeoff Characterization}
To characterize the memory-rate tradeoff for nonuniform file popularity and size, we first propose a lower bound on average rate under uncoded placement.
The lower bound  is  a straightforward extension of the lower bound in {\bf P1} by considering nonuniform file popularity and size, instead of nonuniform file popularity only.

Recall that for nonuniform file popularity and size, $a_{n,l}$ and $M$  are defined in bits in Section~\ref{sec:MCCS_length}. For a given $\abf$, the expression of the lower bound on the average rate $\bar R_\text{lb}(\abf)$ in \eqref{equ:P1_obj} remains unchanged (except that it is in bits).
Since  constraints  \eqref{Constraint2}, \eqref{Constraint_gt0}, and \eqref{non_length_Constraint2} remain the same, we can formulate an optimization problem to minimize $\bar R_\text{lb}(\abf)$ to obtain the lower bound for caching with uncoded placement under nonuniform file popularity and size. The result is given by the following lemma.
\begin{lemma}\label{lemma:lb_length}
For the caching problem with nonuniform file popularity and size, the following optimization problem provides a lower bound on the average rate  for caching with uncoded placement:
\begin{align}
\textrm{\bf P5:}  \min_{\abf}& \; \; \bar R_\text{lb}(\abf)\nn\\
      \textrm{s.t.} & \; \;  \eqref{Constraint2},\eqref{Constraint_gt0}, \; \text{and~} \eqref{non_length_Constraint2}\nn
\end{align}
where  $\bar R_\text{lb}(\abf)$ is given in \eqref{equ:P1_obj}, and   the objective and constraints are all in bits.
 \end{lemma}

Comparing {\bf P4} and {\bf P5}, we  obtain the following result on the optimized MCCS for the two-user case.

\begin{theorem}\label{thm:lb_length}
For  the caching problem of   $N$ files with nonuniform file popularity and size, for $K=2$ users, the minimum average rate for the optimized MCCS in {\bf P4} attains the lower bound given by {\bf P5}.
\end{theorem}
\IEEEproof
See Appendix \ref{Proof:thm:lb_length}.
\endIEEEproof
\begin{Remark}
The tight lower bound shown in Theorem \ref{thm:lb_length} shows the optimality of the optimized  MCCS for $K=2$ users.
It
 enables us to characterize the exact memory-rate tradeoff for $K=2$ users under nonuniform file popularity and size. The optimality of the MCCS also  indicates that there is no loss of optimality by zero-padding. For the general case of $K>2$ users, our numerical studies in Section \ref{sec:num} will show that the gap between the optimized MCCS ({\bf P4}) and the lower bound ({\bf P5}) is very small in general.
\end{Remark}

\section{Numerical Results}\label{sec:num}

In this section, we provide numerically studies on the optimized MCCS and the lower bounds obtained for caching with uncoded placement.
 We first consider files of the same size but with nonuniform  popularity. We study the performance of the optimized MCCS (under the optimal cache placement obtained in Section~\ref{sec:MCCS_placement}), the lower bound in {\bf P1}, and the popularity-first-based lower bound in {\bf P2}. 
For  comparison, we also consider a few existing strategies proposed for the CCS, including i) the optimized CCS~\cite{Deng&Dong:TIT22}, ii) a two-file-group scheme named RLFU-GCC~\cite{Ji&Order:TIT17}, and iii) the mixed caching strategy~\cite{Zhang&Coded:TIT18}. 

\begin{figure}[t]
 \centering
  \includegraphics[width=\columnwidth]{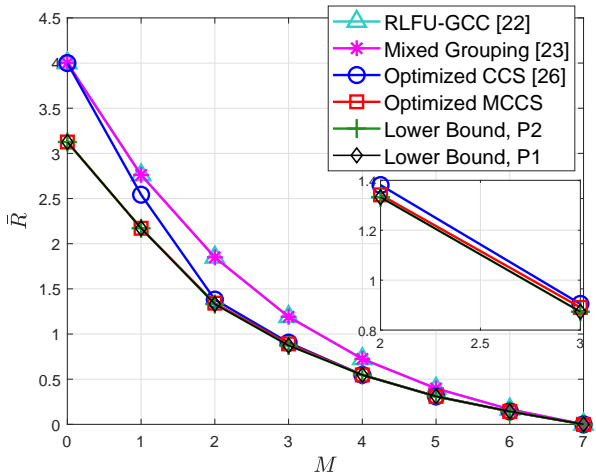}
  \caption{ Average rate $\bar{R}$ vs. cache size $M$  ($N=7$, $K=4$, equal file size, file popularity  Zipf distribution with $\theta=0.56$).}
  \label{fig:perfoamance_1}
  \end{figure}

Let $\bar{R}$ denote the average rate obtained by different schemes or the lower bound. Fig.~\ref{fig:perfoamance_1} shows the average rate $\bar{R}$ vs. $M$  for $N=7$ and  $K=4$.  We generate the file popularities using the Zipf distribution with  $p_{n}={n^{-\theta}}/{\sum_{i=1}^{N}i^{-\theta}}$, where $\theta$ is the Zipf parameter.  We set $\theta=0.56$ (used in  \cite{Daniel&Yu:TIT19,Shanmugam&Golrezaei13IT,zink2009characteristics}).
We see that, among all the caching strategies, the optimized MCCS   results in the lowest average rate for all values of  $M$. The two lower bounds  in {\bf P1} and {\bf P2} are  numerically identical, indicating the optimality of popularity-first placement. Comparing  the optimized MCCS with the lower bounds, we see that the gap between them is very small and only appears at a small range of cache size values  $M \in [2.5,3.5]$. The gap between the average rates of the optimized MCCS and the optimized CCS mainly exists for small cache size $M\in[0,2]$ and shrinks as  $M$ increases. 
 
\begin{table}[t]
\centering
\caption{ The optimal cache placement vectors $\{\abf_n\}$ for the MCCS and the CCS ($M=1$, $N=7$, $K=4$, $\theta=0.56$).}
\resizebox{8.75cm}{!}{
\begin{tabular}{|c|c|c|c|c|c|c|c|c|}
\hline
\multirow{2}{*}{}       & \multirow{2}{*}{$l$} & \multicolumn{7}{c|}{Optimal cache placement vectors for the files}\\ \cline{3-9}
                        &                      & $\abf_1$ & $\abf_2$ & $\abf_3$ & $\abf_4$ & $\abf_5$ & $\abf_6$ & $\abf_7$ \\ \hline\hline
\multirow{5}{*}{CCS}    & $0$                  & 0        & 0        & 0        & 0        & 1.0000   & 1.0000   & 1.0000   \\ 
                        & $1$                  & 0.2500   & 0.2500   & 0.2500   & 0.2500   & 0        & 0        & 0        \\ 
                        & $2$                  & 0        & 0        & 0        & 0        & 0        & 0        & 0        \\ 
                        & $3$                  & 0        & 0        & 0        & 0        & 0        & 0        & 0        \\ 
                        & $4$                  & 0        & 0        & 0        & 0        & 0        & 0        & 0        \\ \hline\hline

\multirow{5}{*}{MCCS}   & $0$                  & 0.4286   & 0.4286   & 0.4286   & 0.4286   & 0.4286   & 0.4286   & 0.4286   \\ 
                        & $1$                  & 0.1429   & 0.1429   & 0.1429   & 0.1429   & 0.1429   & 0.1429   & 0.1429   \\ 
                        & $2$                  & 0        & 0        & 0        & 0        & 0        & 0        & 0        \\ 
                        & $3$                  & 0        & 0        & 0        & 0        & 0        & 0        & 0        \\ 
                        & $4$                  & 0        & 0        & 0        & 0        & 0        & 0        & 0        \\ \hline
\end{tabular}}\label{table:CCS_MCCS_plcement_1}\vspace{2em}
\centering
\caption{ The optimal cache placement vectors $\{\abf_n\}$ for  the MCCS and the CCS ($M=2$, $N=7$, $K=4$, $\theta=0.56$).}
\resizebox{8.75cm}{!}{
\begin{tabular}{|c|c|c|c|c|c|c|c|c|}
\hline
\multirow{2}{*}{}       & \multirow{2}{*}{$l$} & \multicolumn{7}{c|}{Optimal cache placement vectors for the files}\\ \cline{3-9}
                        &                      & $\abf_1$ & $\abf_2$ & $\abf_3$ & $\abf_4$ & $\abf_5$ & $\abf_6$ & $\abf_7$ \\ \hline\hline
\multirow{5}{*}{CCS}    & $0$                  & 0        & 0        & 0        & 0        & 0        & 0        & 0        \\ 
                        & $1$                  & 0.2143   & 0.2143   & 0.2143   & 0.2143   & 0.2143   & 0.2143   & 0.2143   \\ 
                        & $2$                  & 0.0238   & 0.0238   & 0.0238   & 0.0238   & 0.0238   & 0.0238   & 0.0238   \\
                        & $3$                  & 0        & 0        & 0        & 0        & 0        & 0        & 0        \\ 
                        & $4$                  & 0        & 0        & 0        & 0        & 0        & 0        & 0        \\ \hline\hline

\multirow{5}{*}{MCCS}   & $0$                  & 0        & 0        & 0        & 0        & 0        & 0        & 0        \\ 
                        & $1$                  & 0.2143   & 0.2143   & 0.2143   & 0.2143   & 0.2143   & 0.2143   & 0.2143   \\ 
                        & $2$                  & 0.0238   & 0.0238   & 0.0238   & 0.0238   & 0.0238   & 0.0238   & 0.0238   \\
                        & $3$                  & 0        & 0        & 0        & 0        & 0        & 0        & 0        \\ 
                        & $4$                  & 0        & 0        & 0        & 0        & 0        & 0        & 0        \\ \hline

\end{tabular}}\label{table:CCS_MCCS_plcement_2}\vspace{2em}
\centering
\caption{ The optimal cache placement vectors $\{\abf_n\}$ for  the MCCS and the CCS ($M=6$, $N=7$, $K=4$, $\theta=0.56$).}
\resizebox{8.75cm}{!}{
\begin{tabular}{|c|c|c|c|c|c|c|c|c|}
\hline
\multirow{2}{*}{}       & \multirow{2}{*}{$l$} & \multicolumn{7}{c|}{Optimal cache placement vectors for the files}\\ \cline{3-9}
                        &                      & $\abf_1$ & $\abf_2$ & $\abf_3$ & $\abf_4$ & $\abf_5$ & $\abf_6$ & $\abf_7$ \\ \hline\hline
\multirow{5}{*}{CCS}   & $0$                  & 0        & 0        & 0        & 0        & 0        & 0        & 0        \\ 
                        & $1$                  & 0        & 0        & 0        & 0        & 0        & 0        & 0        \\ 
                        & $2$                  & 0        & 0        & 0        & 0        & 0        & 0        & 0        \\ 
                        & $3$                  & 0.4286   & 0.4286   & 0.4286   & 0.4286   & 0.4286   & 0.4286   & 0.4286   \\ 
                        & $4$                  & 0.1429   & 0.1429   & 0.1429   & 0.1429   & 0.1429   & 0.1429   & 0.1429   \\ \hline\hline

\multirow{5}{*}{MCCS}   & $0$                  & 0        & 0        & 0        & 0        & 0        & 0        & 0        \\ 
                        & $1$                  & 0        & 0        & 0        & 0        & 0        & 0        & 0        \\ 
                        & $2$                  & 0        & 0        & 0        & 0        & 0        & 0        & 0        \\ 
                        & $3$                  & 0.4286   & 0.4286   & 0.4286   & 0.4286   & 0.4286   & 0.4286   & 0.4286   \\ 
                        & $4$                  & 0.1429   & 0.1429   & 0.1429   & 0.1429   & 0.1429   & 0.1429   & 0.1429   \\ \hline

\end{tabular}}\label{table:CCS_MCCS_plcement_3}
\end{table}

As discussed in Section \ref{sec:MCCS_structure}, although the candidate solutions of the optimal cache placement for the MCCS and the CCS are the same,  the optimal placements  may be different for the two schemes. To see this difference, for the same setting considered in Fig.~\ref{fig:perfoamance_1}, we show the optimal $\{\abf_n\}$ for the two schemes for $M=1,2,6$ in Tables~\ref{table:CCS_MCCS_plcement_1}, \ref{table:CCS_MCCS_plcement_2}, and \ref{table:CCS_MCCS_plcement_3}, respectively, representing small, moderate, and large cache size. For a small cache size ($M=1$), the optimal placements for the MCCS and the CCS in Table \ref{table:CCS_MCCS_plcement_1} are different. For the MCCS, all files have the identical placement, where each file is partitioned into subfiles of two 
sizes, with one stored at the server ($a_{n,0}$) and the rest in each user's local cache. In contrast, for the CCS, files $\{W_5,W_6,W_7\}$ are solely stored at the server, and files $\{W_1,\cdots, W_4\}$ are stored in each user's local cache. This difference on the placement is the main cause of the performance gap between the MCCS and the CCS in Fig.~\ref{fig:perfoamance_1}.  For moderate to large cache size ($M=2,6$),  Tables~\ref{table:CCS_MCCS_plcement_2} and \ref{table:CCS_MCCS_plcement_3} show that the optimal cache placements are the same for the MCCS and the CCS. However, we  see from Fig.~\ref{fig:perfoamance_1}  that for $M=2$, there is a small observable gap between the average rates of the two schemes, with that of the MCCS being lower; and for $M=6$, the average rates of the two are nearly identical. The explanation for this trend is that  there exist more redundant messages  for $M=2$ with the placement in Table~\ref{table:CCS_MCCS_plcement_2} than those for $M=6$ with the placement in Table~\ref{table:CCS_MCCS_plcement_3}. To elaborate more on this, note that for given demand $\dbf$,  the number of redundant groups in cache subgroup $\Ac^{l+1}$ is $\binom{K-\tilde N(\dbf)}{l}$, which decreases with $l$. They determine the number of redundant messages. The indices of  the nonzero elements in $\abf_n$ are $l=1,2$ for $M=2$, and $l=3,4$ for $M=6$.   As a result, for $M=6$, there are only a very small number of  redundant messages that are removed by the MCCS. Thus, the performance of the MCCS and the CCS are almost identical. Finally, the larger improvement of the MCCS over the CCS (and the MCCS almost attains the lower bounds)\  for $M\in [0,2]$  indicates that,  at a small cache size, coded caching is more sensitive to the cache placement to achieve the largest  caching gain.

\begin{figure}[t]
   \centering
  \includegraphics[width=\columnwidth]{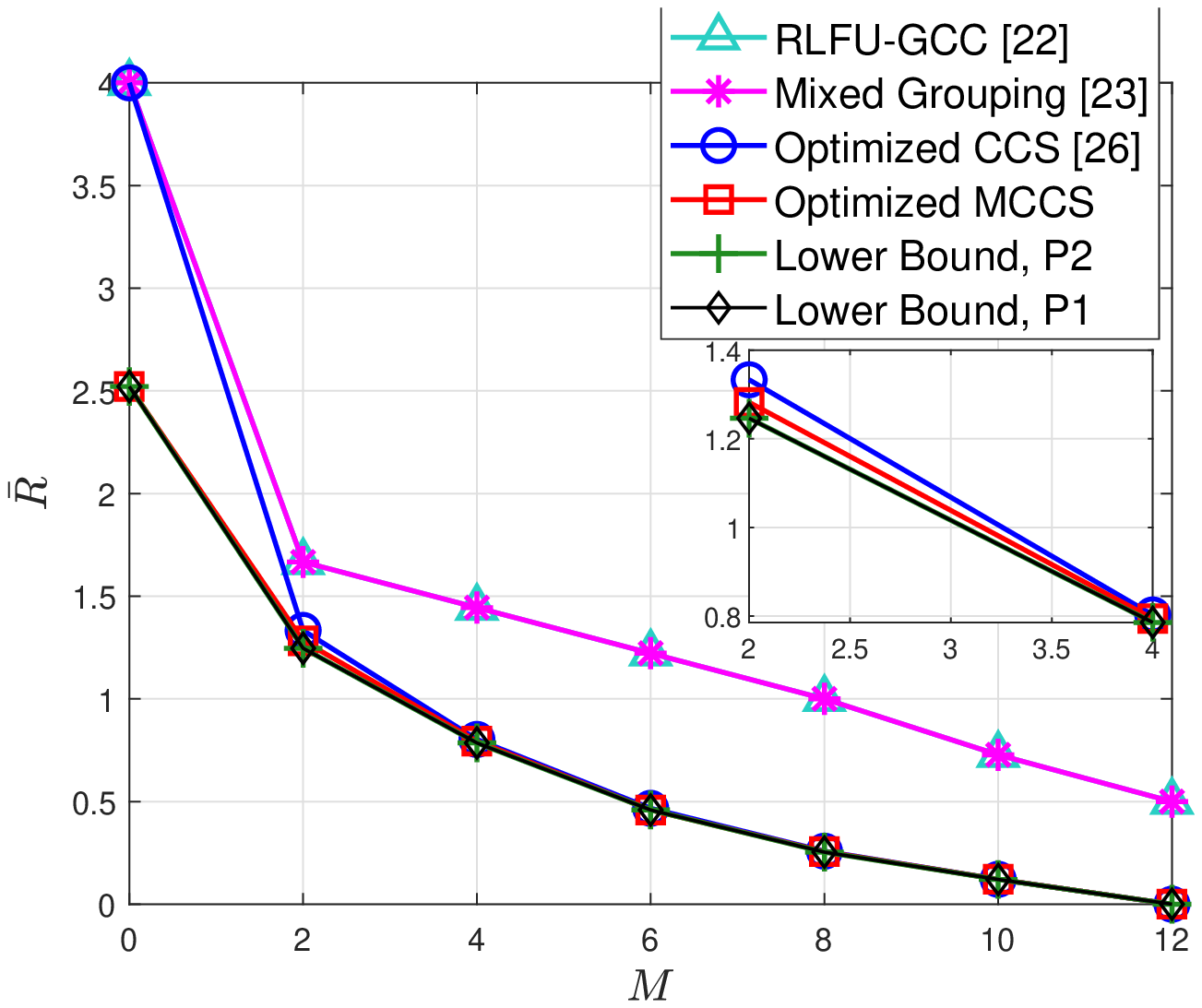}
  \caption{Average rate $\bar{R}$ vs. cache size $M$ ($N=12$, $K=4$, equal file size, file popularity distribution: step function).}
  \label{fig:perfoamance_2}\vspace*{1.5em}
  \centering
  \includegraphics[width=\columnwidth]{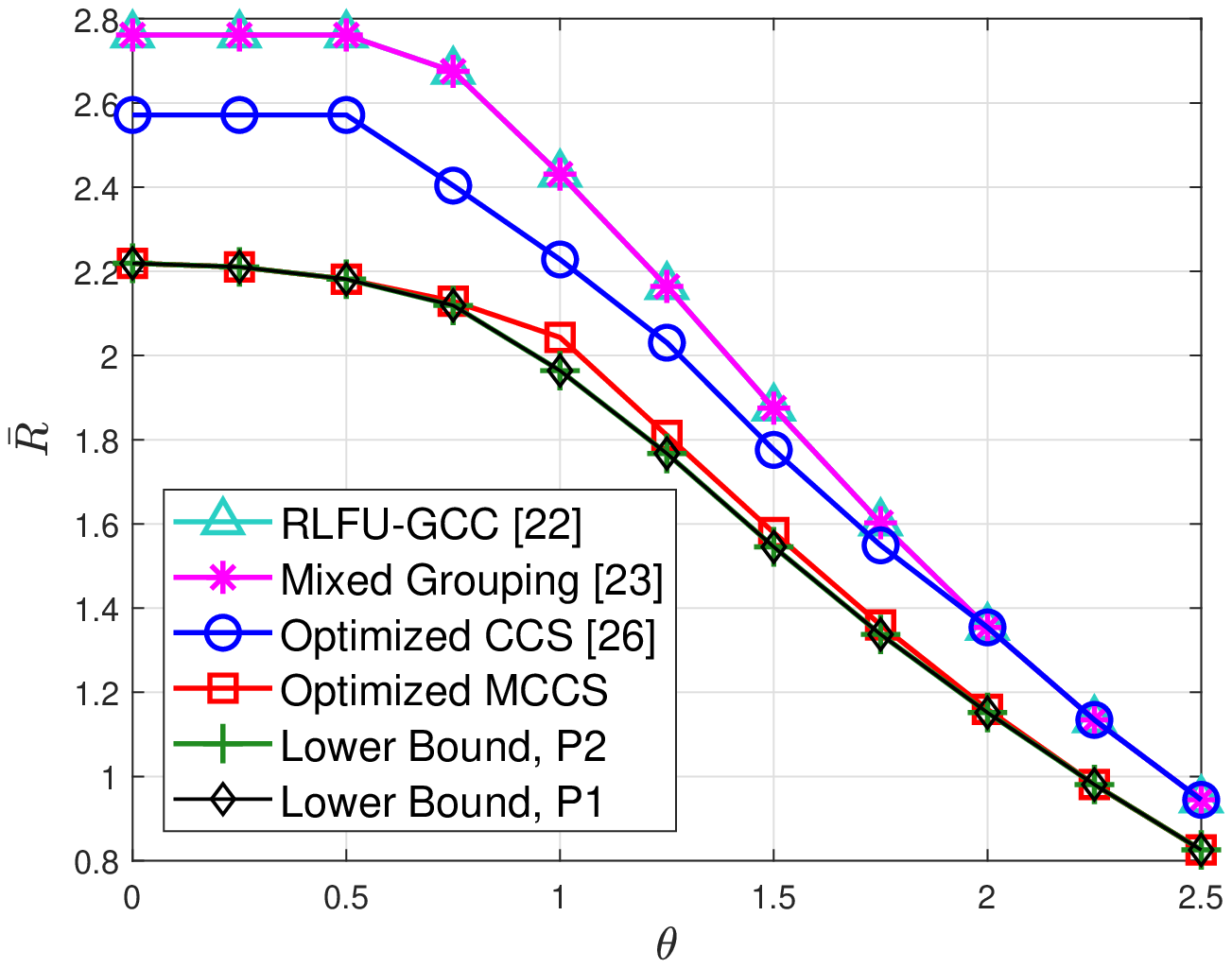}
  \caption{Average rate $\bar{R}$ vs. Zipf parameter $\theta$ ($N=7$, $K=4$,  $M=1$, equal file size).}
  \label{fig:perfoamance_3}
\end{figure}

To evaluate the performance with other file popularity distribution, we consider the case studied in \cite{Zhang&Coded:TIT18} with $N=12$, $K=5$, and a  step-function for file popularity distribution: $p_1=7/12$, $p_n=1/18$, $n=2,\ldots, 7$, and $p_n=1/60$, $n=8,\ldots, 12$.   Fig.~\ref{fig:perfoamance_2} shows the average rate $\bar{R}$ vs. $M$ by different caching schemes and the lower bounds.
Similar to Fig.~\ref{fig:perfoamance_1}, for all values of  $M$, the optimized MCCS  achieves the lowest $\bar{R}$ among all the strategies, which  is very close to the lower bounds. The two lower bounds in {\bf P1} and {\bf P2} are equal for different values of $M$, with the only exception for $M=2$, where $\bar R$ for {\bf P1} is $10^{-4}$ smaller than that of {\bf P2}.  The gap between the MCCS and the CCS again only exists for small values of $M$.   To show the performance at different levels of popularity distribution,  we show in Fig. \ref{fig:perfoamance_3} the average rate  $\bar{R}$  vs. Zipf  parameter $\theta$. We set $N=7$, $K=4$. We choose a small cache size $M=1$ to   show clearly the performance gap between the caching schemes and lower bounds. The optimized MCCS always performs the best among all the caching strategies for any $\theta$. The lower bound in {\bf P1} and the popularity-first-based lower bound in {\bf P2} are numerically identical.
Also, we observe that the gap between  the average rates of the optimized MCCS and the lower bounds only exists at a moderate range of $\theta$  and is very  small in general. In contrast, the gap between the MCCS and the CCS is obvious at  all values of $\theta$. This demonstrates the advantage of the MCCS over other caching schemes at a small value of $M$.
\begin{table}[t]
\centering
\caption{ file grouping structures of the optimal cache placement $\{\abf_n\}$ for  the MCCS   ($N=9$, $K=4$, $\theta=1.2$).}
\resizebox{8.75cm}{!}{
\begin{tabular}{|c|c|c|c|c|c|c|c|c|c|c|}
\hline
\multirow{2}{*}{$M$} & \multirow{2}{*}{$l$} & \multicolumn{9}{c|}{Optimal cache placement vectors for the files}                               \\ \cline{3-11}
                     &                      & $\abf_1$ & $\abf_2$ & $\abf_3$ & $\abf_4$ & $\abf_5$ & $\abf_6$ & $\abf_7$ & $\abf_8$ & $\abf_9$ \\ \hline
\multirow{5}{*}{$3$} & $0$                  & 0        & 0        & 0        & 0        & 1.000    & 1.000   & 1.000      & 1.000    & 1.000          \\
                     & $1$                  & 0        & 0        & 0        & 0        & 0        & 0        & 0         & 0        & 0        \\
                     & $2$                  & 0        & 0        & 0        & 0        & 0        & 0        & 0         & 0        & 0        \\
                     & $3$                  & 0.250    & 0.250    & 0.250    & 0.250    & 0        & 0        & 0        & 0        & 0        \\
                     & $4$                  & 0        & 0        & 0        & 0        & 0        & 0        & 0        & 0        & 0         \\ \hline\hline
\multirow{5}{*}{$4$} & $0$                  & 0        & 0        & 0        & 0        & 0        & 0.667   & 1.000     & 1.000   & 1.000          \\
                     & $1$                  & 0        & 0        & 0        & 0        & 0        & 0        & 0        & 0        & 0          \\
                     & $2$                  & 0        & 0        & 0        & 0        & 0        & 0        & 0        & 0        & 0          \\
                     & $3$                  & 0.250    & 0.250    & 0.250    & 0.250    & 0.250    & 0.083    & 0         & 0        & 0         \\
                     & $4$                  & 0        & 0        & 0        & 0        & 0        & 0        & 0       & 0        & 0        \\ \hline\hline
\multirow{5}{*}{$7$} & $0$                  & 0        & 0        & 0        & 0        & 0        & 0        & 0        & 0        & 0         \\
                     & $1$                  & 0        & 0        & 0        & 0        & 0        & 0        & 0        & 0        & 0          \\
                     & $2$                  & 0        & 0        & 0        & 0        & 0        & 0        & 0        & 0        & 0          \\
                     & $3$                  & 0.222    & 0.222    & 0.222    & 0.222    & 0.222    & 0.222    & 0.222    & 0.222    & 0.222         \\
                     & $4$                  & 0.111    & 0.111    & 0.111    & 0.111    & 0.111    & 0.111    & 0.111    & 0.111    & 0.111         \\ \hline
\end{tabular}}\label{tbl:file_grouping}
\end{table}

We now verify   the structure of the optimal cache placement  for the  MCCS described in Section \ref{sec:MCCS_structure}.
We  generate file popularity using Zipf distribution with $\theta=1.2$. We obtain the optimal placement solution $\{\abf_n\}$ using our proposed algorithm and verify that it matches the optimal solution obtained by solving \textbf{P0} numerically. As an example, for $N=9$ and $K=4$, Table~\ref{tbl:file_grouping} shows  the optimal $\{\abf_n\}$ that is obtained  by solving \textbf{P0} numerically, for $M=3,4,7$. For $M=3$, we see that  there are two file groups $\{W_1,\ldots,W_4\}$ and $\{W_5,\ldots,W_9\}$ under the optimal placement. This structure matches  Case 1 in Section \ref{sec:nonZeroSecond}, where the cache is entirely allocated to the first file group with the most popular files, and the  files in the second file group are only stored at the server ($a_{5,0}=\ldots =a_{9,0}=1$). The optimal $\abf_n$'s for the first group are identical with  only one nonzero element. This means those files are partitioned into subfiles of the same size and are stored at users' local caches. With a small cache size, this placement result is intuitive: only a few popular files are cached, and the rest remain in the server; thus,  the optimal cache placement results in two file groups.
For $M=4$, a different  cache placement strategy is shown, where the files are divided into three file groups. The optimal $\{\abf_n\}$ is as described in  Section \ref{sec:threeGroups} for the three-file-group case:  no cache is allocated to the third file group $\{W_7,W_8,W_9\}$, and a portion of the file is cached for $W_6$ in the second file group;  for the first file group, files are partitioned into subfiles of a single size and are all stored at different users. For $M=7$, the optimal placement has only a single file group,  where all the files have the same placement as discussed in Section \ref{sec:oneGroup}. From Table~\ref{tbl:file_grouping}, we see that  the file popularity differences are more critical for the placement when the case size $M$ is limited (relative to the total files in the database).
As  
$M$ becomes large, all the files tend to have the same placement  into the user caches. 

\begin{figure}[t]
 \centering
  \includegraphics[width=\columnwidth]{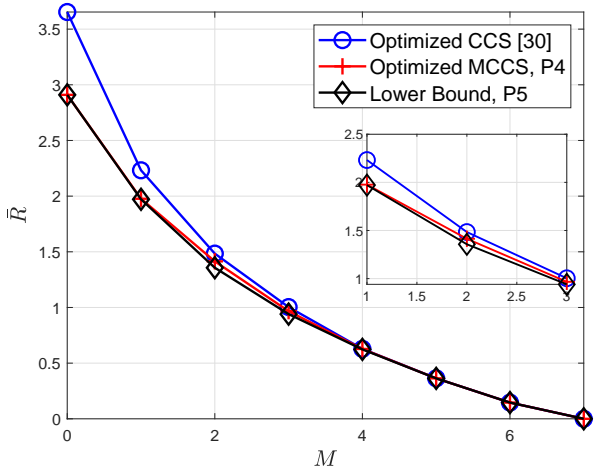}
  \caption{Average rate $\bar{R}$ vs. cache size $M$ (kbits) ($N=7$, $K=4$, file popularity distribution: $\pbf=[0.0888,0.0968,0.1072,0.1215,0.2640,0.1427,0.1791]$, file sizes: $[F_1,\ldots,F_N]=[9/6, 8/6, 7/6,6/6,5/6, 4/6, 3/6]$ kbits).}
  \label{fig:nonuni_popu_length_1}
\end{figure}

Finally,  we consider the scenario of nonuniform file popularity and size. We generate the  file popularity using Zipf distribution with $\theta=0.56$, which gives $\pbf=[0.0888,0.0968,0.1072,0.1215,0.2640,0.1427,0.1791]$. The file sizes  are set as $[F_1,\ldots,F_N]=[9/6, 8/6, 7/6, 6/6, 5/6, 4/6, 3/6]$ kbits. The file size and popularity
combinations  are chosen similar to those  used in \cite{Daniel&Yu:TIT19}, which  simulate a practical scenario where file popularity and size are relatively uncorrelated. In Fig. \ref{fig:nonuni_popu_length_1}, we compare the  optimized MCCS in {\bf P4}, the lower bound in {\bf P5}, and the optimized CCS~\cite{Daniel&Yu:TIT19}. The gap between the optimized MCCS and the lower bound only exists for $M\in[1,3] $ and is very small.
Moreover, the optimized MCCS outperforms the optimized CCS. The gap between the two again is obvious at small values of $M$, and it reduces to zero as $M$ becomes large. 

\section{Conclusion}\label{sec:conclusion}
In this paper, for a caching system with nonuniform file popularity, we characterized the memory-rate tradeoff for caching with uncoded placement. We focused on the\ MCCS with cache placement optimized in the class of  popularity-first placement for average rate minimization. We then provided a general lower bound and a popularity-first-based lower bound for caching with uncoded placement. For $K=2$ users, the two lower bounds are shown to be identical, and the optimized MCCS attains the bounds, providing the exact memory-rate tradeoff. For   $K>2$ users with distinct  requests, the optimized MCCS attains the popularity-first-based lower bound. Additionally, the results  in these two regions reveal  the following results for the MCCS unavailable in the literature: The  popularity-first placement is optimal, and zero-padding used in coded delivery incurs no loss of optimality. For $K>2$ users with redundant requests, our analysis showed that a gap might exist between the optimized MCCS and the lower bounds due to zero-padding. However, numerical results show that such loss only exists in some limited cases and is very small in general. 

We next  characterized the optimal solution structure of the  popularity-first cache placement for the MCCS under nonuniform file popularity. It was shown to have a simple file-grouping structure of at most three file groups, depending on the relative cache size to the database size. We obtained the closed-form placement solution for each candidate structure,
which enabled us to compute the optimal solution through a simple  algorithm. Finally, we extended our study of memory-rate tradeoff to the case where files are nonuniform  in both popularity and size. We showed that the optimized MCCS attains the lower bound for $K=2$ users and characterizes the exact memory-rate tradeoff.
Numerical results again showed that, for general settings, the gap between the optimized MCCS and the lower bound exists in limited cases and is very small.

\appendices
\section{Proof of Lemma \ref{lemma_bnd_2}}\label{Proof:lemma_bnd_2}
\IEEEproof
The proof directly follows  the proof of  \cite[Theorem 2]{Sahraei2019TheOptimal} with a slight modification. In the proof of  \cite[Theorem 2]{Sahraei2019TheOptimal}, by a genie-based method, it is shown that the delivery rate for a  distinct file set ${\Dc}$ satisfies
\begin{align} \label{app:R_D}
 R(\Dc)\ge\max_{\pi:\Ic_{|\!\Dc|}\rightarrow\Dc} \sum_{l=0}^{K}\sum_{i=1}^{|\Dc|}\binom{K-i}{l}\frac{\check a_{\pi(i),l}}{\binom{K}{l}}
\end{align} where $\check a_{\pi(i),l}$ is the
 total number of bits from file $\pi(i)\in\Dc$ cached by exactly $l$ users. In our definition of $a_{\pi(i),l}$, subscript $l$ refers to the user subset size $|\Sc|=l$ in a cache subgroup $\Ac^l$, which is  identical to that in $\check a_{\pi(i),l}$.  Following the definitions of  $\check a_{\pi(i),l}$ and $a_{\pi(i),l}$, we have $a_{\pi(i),l}={\check a_{\pi(i),l}}/{\binom{K}{l}}$, since there are $\binom{K}{l}$ user subsets in $\Ac^l$. Thus, for  given $\abf$, we can equivalently express \eqref{app:R_D} as
\begin{align}
 R(\Dc)\ge\!\max_{\pi:\Ic_{|\!\Dc|}\rightarrow\Dc} \sum_{l=0}^{K}\sum_{i=1}^{|\Dc|}\binom{K-i}{l}a_{\pi(i),l}\triangleq R_\text{lb}(\Dc;\abf).
\end{align}
Using the above expression, by averaging $R_\text{lb}(\Dc;\abf)$  over all possible $\Dc\subseteq\Nc$, we obtain the general lower bound  $\bar R_\text{lb}(\abf)$  the average rate w.r.t $\abf$ in \eqref{equ:P1_obj}. The final lower bound on average rate is obtained by optimizing $\abf$ to minimize $\bar R_\text{lb}(\abf)$, which is shown in {\bf P1}.
\endIEEEproof

\section{Proof of Lemma \ref{lemma:lbpf}}\label{Proof:lemma_lbpf}
\IEEEproof
The popularity-first-based lower bound is essentially a simplification of  {\bf P1} in Lemma \ref{lemma_bnd_2}, by restricting to the set of popularity-first placement vectors: $\abf\in\Qc$. We need to show that $R_\text{lb}({\Dc};\abf)$ in  \eqref{R_lblb} can be simplified into \eqref{equ:bnd_5} for $\abf\in\Qc$. For $\forall\Dc\subseteq\Nc$, since $\phi(\cdot)$ is such that  $p_{\phi(1)} \ge \cdots \ge p_{\phi(|\Dc|)}$, by the definition of popularity-first placement in \eqref{equ:popu_fir}, for $\abf\in\Qc$,  we have
\begin{align}
a_{\phi(i),l}\ge a_{\phi(i+1),l},\quad l\in\Kc, \; i=1,\ldots, |\Dc|-1.
\end{align}
Since $\binom{K-i}{l}$ is a decreasing function of $i$, for $\abf\in\Qc$,
we have
\begin{align*}
\max_{\pi:\Ic_{|\!\Dc|}\rightarrow\Dc}& \sum_{l=0}^{K} \sum_{i=1}^{|\Dc|}\!\binom{K-i}{l}a_{\pi(i),l} =\sum_{l=0}^{K}\sum_{i=1}^{|\Dc|}\!\binom{K-i}{l}a_{\phi(i),l}.
\end{align*}
 Thus, we remove the max operation in \eqref{R_lblb} to arrive at the simplified expression in \eqref{equ:bnd_5}, for $\abf\in\Qc$.
\endIEEEproof

\section{Proof of Theorem \ref{thm:lbequ}}\label{Proof:thm:lbequ}
\IEEEproof
For  {\bf P1}, consider a feasible cache placement vector $\hat\abf$ and any $l_o\in\Kc$. Define $\varphi: \Ic_{|\Nc|}\rightarrow\Nc$ as a bijective map for $\hat{\abf}$ such that $\hat a_{\varphi(1),l_o}\ge\ldots\ge \hat a_{\varphi(N),l_o}$.
Note that $\varphi(\cdot)$ depends on $\hat{\abf}$ and  $l_o$.

Assume   $p_{\varphi(i_o)}< p_{\varphi(i_o+1)}$, for some $i_o\in\Ic_{|\Nc|}\backslash\{N\}$. We construct another feasible cache placement
vector $\tilde\abf$ using $\hat{\abf}$ by switching the values of $\hat a_{\varphi(i_o),l_o}$ and $\hat a_{\varphi(i_o+1),l_o}$. Specifically,\\ i) for $l \in \Kc$, we have
\begin{align}\label{equ:tilde_a_l}
\hspace*{-8em} \begin{cases}
\tilde a_{\varphi(i_o),l_o} = \hat a_{\varphi(i_o+1),l_o}\\
\tilde a_{\varphi(i_o+1),l_o} = \hat a_{\varphi(i_o),l_o}\\
\tilde a_{\varphi(i),l}=\hat a_{\varphi(i),l},\quad i\neq i_o, i \in \Ic_{|\Nc|};
\end{cases}
\end{align}
ii) for $l=0$, by  \eqref{equ:tilde_a_l} and  file partition constraint  \eqref{Constraint1.1}, we have
\begin{align}\label{equ:a_n0}
\begin{cases}
\tilde a_{\varphi(i_o),0}=1\!-\!\!\sum_{l\in \Kc\backslash\{l_o\}}\!\!\binom{K}{l}\hat a_{\varphi(i_o),l}-\binom{K}{l_o}\hat a_{\varphi(i_o+1),l_o}\\
\tilde a_{\varphi(i_o+1),0}=1\!-\!\!\sum_{l\in \Kc\backslash\{l_o\}}\!\!\binom{K}{l}\hat a_{\varphi(i_o+1),l}-\binom{K}{l_o}\hat a_{\varphi(i_o),l_o}\\
\tilde a_{\varphi(i),0}=\hat a_{\varphi(i),0}, \quad i\neq i_o,i_o+1, \; i \in \Ic_{|\Nc|}.
\end{cases}
\end{align}
From \eqref{equ:tilde_a_l}, we have
\begin{align}
&\tilde{a}_{\varphi(1),l_o}\ge \cdots \ge \tilde{a}_{\varphi(i_o),l_o}, \;\;  \tilde{a}_{\varphi(i_o+1),l_o}\ge \cdots \ge \tilde{a}_{\varphi(N),l_o}, \nn\\
&\tilde{a}_{\varphi(i_o+1),l_o}\ge\tilde{a}_{\varphi(i_o),l_o}. \label{equ:tilde_a_io}
\end{align}

From \eqref{equ:a_n0} and  \eqref{Constraint1.1}, we   conclude that
\begin{align}
&\!\!\hat a_{\varphi(i_o),0}+\hat a_{\varphi(i_o+1),0}=\tilde a_{\varphi(i_o),0}+\tilde a_{\varphi(i_o+1),0}\label{equ:an0_1}\\
&\!\!\hat a_{\varphi(i_o),0}-\tilde a_{\varphi(i_o),0}=\binom{K}{l_o}(\hat a_{\varphi(i_o+1),l_o}-\hat a_{\varphi(i_o),l_o})\label{equ:an0_2}\\
&\!\!\hat a_{\varphi(i_o+1),0}\!-\!\tilde a_{\varphi(i_o+1),0}\!=\!\binom{K}{l_o}(\hat a_{\varphi(i_o),l_o}\!-\!\hat a_{\varphi(i_o+1),l_o}).\label{equ:an0_3}
\end{align}
Now, we  show that $\bar R_\text{lb}(\hat\abf)\ge\ \bar R_\text{lb}(\tilde\abf)$.

For $K=2$, we have
$\tilde{N}(\dbf)=|\Dc|\le2$. Define  $\xi:\Ic_{|\!\Dc|}\rightarrow\Dc$ as a bijective map for $\abf$ such that $a_{\xi(1),1}\ge a_{\xi(|\Dc|),1}$. Note that  $\xi(\cdot)$ depends on  $\abf$. Then, $R_\text{lb}(\Dc;\abf)$ in \eqref{R_lblb} is given by\\[-1.5em]
\begin{align}\label{equ:R_lb_K2}
R_\text{lb}(\Dc;\abf)
&=\max_{\pi:\Ic_{|\!\Dc|}\rightarrow\Dc}\Big\{\sum_{i=1}^{|\Dc|}a_{\pi(i),0}+a_{\pi(1),1}\Big\} \nn\\
&=\sum_{i=1}^{|\Dc|}a_{\xi(i),0}+a_{\xi(1),1}
\end{align}

Given $\hat{\abf}$, the set  $\Dc$ of distinct file indices in demand vector $\dbf$ can be categorized into the following four types:

\begin{enumerate}
\item $\widetilde{\Dc}_{1,j}\triangleq\{\Dc\subseteq\Nc:\varphi(i_o)\in\Dc,\varphi(i_o+1)\notin\Dc,\varphi(i_o)=\xi(j)\}$, for $j=1,2$.\footnote{Set $\widetilde{\Dc}_{1,j}$ corresponds to the case where file  $\varphi(i_o)$ is requested and $a_{\varphi(i_o),1}$ is ranked the $j$th in $\xi(\cdot)$.}
\item $\widetilde{\Dc}_{2,j}\triangleq\{\Dc\subseteq\Nc:\varphi(i_o+1)\in\Dc,\varphi(i_o)\notin\Dc,\varphi(i_o+1)=\xi(j)\}$, for $j=1,2$.
\item $\widetilde{\Dc}_{3}\triangleq\{\;\{\varphi(i_o), \varphi(i_o+1)\} \;\}$.
\item $\widetilde{\Dc}_4\triangleq\{\Dc\subseteq\Nc \backslash\{\varphi(i_o), \varphi(i_o+1)\} \; \}$.
\end{enumerate}

Note that for any  $\Dc$, its type is the same for  $\hat\abf$ and $\tilde\abf$. To see this, consider  $\Dc=\{\varphi(i_o),n'\}$, where $n'\in\Nc\backslash\{\varphi(i_o),\varphi(i_o+1)\}$. Assume $\hat{a}_{\varphi(i_o),1}\ge\hat{a}_{n',1}$. Then for $\hat{\abf}$, we have $\xi(1)=\varphi(i_o)$, and $\Dc\in\widetilde{\Dc}_{1,1}$. For  $\tilde\abf$, from \eqref{equ:tilde_a_io}, we also have $\tilde{a}_{\varphi(i_o),1}\ge\tilde{a}_{n',1}$. Thus, for the mapping $\xi(\cdot)$ for  $\tilde\abf$, we have  $\xi(1)=\varphi(i_o)$,  and in this case, we again have $\Dc\in\widetilde{\Dc}_{1,1}$. All other types of $\Dc$ can be verified using the similar argument.

Based on the above four categories of $\Dc$, we rewrite $\bar{R}_\text{lb}(\abf)$ in \eqref{equ:P1_obj} as
\begin{align*}
\bar R_\text{lb}(\abf)=&\sum_{i=1}^2\sum_{j=1}^{2}\sum_{\Dc\in\widetilde{\Dc}_{i,j}}\sum_{\dbf\in\Tc(\Dc)}p_{d_1}p_{d_2}R_\text{lb}(\Dc;\abf)\nn\\
&+\!\!\sum_{\Dc\in \widetilde{\Dc}_{3} \cup \widetilde{\Dc}_{4}}\sum_{\dbf\in\Tc(\Dc)}p_{d_1}p_{d_2}R_\text{lb}(\Dc;\abf).
\end{align*}
Following the above, we have
\begin{align}\label{equ:lb_diff}
&\bar R_\text{lb}(\hat\abf)-\bar R_\text{lb}(\tilde\abf)\nn\\
&=\sum_{i=1}^2\sum_{j=1}^{2}\sum_{\Dc\in\widetilde{\Dc}_{i,j}}\!\sum_{\dbf\in\Tc(\Dc)}p_{d_1}p_{d_2}(R_\text{lb}(\Dc;\hat\abf)-R_\text{lb}(\Dc;\tilde \abf))\nn\\
&\quad+\!\!\sum_{\Dc\in \widetilde{\Dc}_{3} \cup \widetilde{\Dc}_{4}}\!\sum_{\dbf\in\Tc(\Dc)}p_{d_1}p_{d_2}(R_\text{lb}(\Dc;\hat\abf)-R_\text{lb}(\Dc;\tilde \abf)).
\end{align}

We now evaluate the differences between $R_\text{lb}(\Dc;\hat\abf)$ and $R_\text{lb}(\Dc;\tilde\abf)$, for $l_o=1,2$.

\emph{\textbf{Case 1: $l_o=1$.}}
 We  express $R_\text{lb}(\Dc;\hat\abf)$ in  \eqref{equ:R_lb_K2} based on the types of $\Dc$.
If $\Dc\in\widetilde{\Dc}_{1,1}$, then $\varphi(i_o)=\xi(1)$, and we have
\begin{align}\label{equ:lb_a_hat_D11_0}
\hspace*{-.8em}R_\text{lb}(\Dc;\hat\abf)=
\begin{cases}
\hat a_{\varphi(i_o),0}+\hat a_{\varphi(i_o),1},\quad |\Dc|=1\\
\hat a_{\varphi(i_o),0}+\hat a_{\xi(2),0}+\hat a_{\varphi(i_o),1},\quad |\Dc|=2,
\end{cases}
\end{align}
or more compactly, we can express \eqref{equ:lb_a_hat_D11_0} as follows
\begin{align}\label{equ:lb_a_hat_D11}
R_\text{lb}(\Dc;\hat\abf)=\hat a_{\varphi(i_o),0}+s\cdot\hat a_{\xi(2),0}+\hat a_{\varphi(i_o),1}, \; \Dc\in\widetilde{\Dc}_{1,1}
\end{align}
where $s\in \{0,1\}$ is an indicator defined by $s=\{0: \text{if}~ |\Dc|=1; 1: \text{if}~|\Dc|=2\}$. Similarly, for any other types of $\Dc$, we can always rewrite  $R_\text{lb}(\Dc;\hat\abf)$  in  \eqref{equ:R_lb_K2} as in  \eqref{equ:lb_a_hat_D11} by replacing $\xi(1)$ and $\xi(2)$ with $\varphi(i_o)$ and $\varphi(i_o+1)$, given as follows
\begin{align}\label{equ:lb_a_hat}
R_\text{lb}(\Dc;\hat\abf)\!=\!\begin{cases}
\hat a_{\varphi(i_o),0}+s\cdot\hat a_{\xi(2),0}+\hat a_{\varphi(i_o),1}, &\!\! \Dc\in\widetilde{\Dc}_{1,1} \\
\hat a_{\xi(1),0}+\hat a_{\varphi(i_o),0}+\hat a_{\xi(1),1},&\!\!\Dc\in\widetilde{\Dc}_{1,2}\\
\hat a_{\varphi(i_o+1),0}+s\cdot\hat a_{\xi(2),0}+\hat a_{\varphi(i_o+1),1},&\!\!\Dc\in\widetilde{\Dc}_{2,1}\\
\hat a_{\xi(1),0}+\hat a_{\varphi(i_o+1),0}+\hat a_{\xi(1),1},&\!\!\Dc\in\widetilde{\Dc}_{2,2}\\
\hat a_{\varphi(i_o),0}+\hat a_{\varphi(i_o+1),0}+\hat a_{\varphi(i_o),1}, &\!\! \Dc\in\widetilde{\Dc}_3\\
\hat{a}_{\xi(1),0}+s\cdot \hat{a}_{\xi(2),0}+\hat{a}_{\xi(1),1},&\!\!\Dc\in\widetilde{\Dc}_4
\end{cases}
\end{align}
where the second and fourth cases are only for $|\Dc|=2$.

Similar to  $R_\text{lb}(\Dc;\hat\abf)$ in \eqref{equ:lb_a_hat},  we can rewrite   $R_\text{lb}(\Dc;\tilde{\abf})$ in \eqref{equ:R_lb_K2}   as follows
 \begin{align}\label{equ:lb_a_tilde}
\!R_\text{lb}(\Dc;\tilde\abf)\!=\!\begin{cases}
\tilde{a}_{\varphi(i_o),0}+s\cdot\tilde{a}_{\xi(2),0}+\tilde{a}_{\varphi(i_o),1},&\!\!\Dc\in\widetilde{\Dc}_{1,1}\\
\tilde{a}_{\xi(1),0}+\tilde{a}_{\varphi(i_o),0}+\tilde{a}_{\xi(1),1},&\!\!\Dc\in\widetilde{\Dc}_{1,2}\\
\tilde{a}_{\varphi(i_o+1),0}+s\cdot\tilde{a}_{\xi(2),0}+\tilde{a}_{\varphi(i_o+1),1},&\!\!\Dc\in\widetilde{\Dc}_{2,1}\\
\tilde{a}_{\xi(1),0}+\tilde{a}_{\varphi(i_o+1),0}+\tilde{a}_{\xi(1),1},&\!\!\Dc\in\widetilde{\Dc}_{2,2}\\
\tilde{a}_{\varphi(i_o),0}+\tilde{a}_{\varphi(i_o+1),0}+\tilde{a}_{\varphi(i_o+1),1},&\!\!\Dc\in\widetilde{\Dc}_3\\
\tilde{a}_{\xi(1),0}+s\cdot \tilde{a}_{\xi(2),0}+\tilde{a}_{\xi(1),1}\, &\!\!Dc\in\widetilde{\Dc}_4.
\end{cases}
\end{align}
Comparing \eqref{equ:lb_a_hat} and \eqref{equ:lb_a_tilde}, we note that the only difference   is the case of $\Dc \in \widetilde{\Dc}_3$, where $\hat{a}_{\xi(1),1}=\hat{a}_{\varphi(i_o),1}$, while  $\tilde{a}_{\xi(1),1}=\tilde{a}_{\varphi(i_o+1),1}$ by \eqref{equ:tilde_a_io}.

For $K=2$, $l_o=1$, \eqref{equ:an0_2} and \eqref{equ:an0_3} are respectively given by
\begin{align}
&\hat a_{\varphi(i_o),0}-\tilde{a}_{\varphi(i_o),0}=2(\hat a_{\varphi(i_o+1),1}-\hat a_{\varphi(i_o),1}),\label{equ:an0_2_i}\\
&\hat a_{\varphi(i_o+1),0}-\tilde{a}_{\varphi(i_o+1),0}\!=2(\hat a_{\varphi(i_o),1}-\hat a_{\varphi(i_o+1),1}).\label{equ:an0_3_i}
\end{align}

Based on \eqref{equ:lb_a_hat}--\eqref{equ:an0_3_i}, we now compute $R_\text{lb}(\Dc;\hat \abf)-R_\text{lb}(\Dc;\tilde \abf)$ for different types of $\Dc$.
For $\Dc\in\widetilde{\Dc}_{1,1}$, we have
\begin{align}
R_\text{lb}(\Dc;\hat\abf)\!-\!R_\text{lb}(\Dc;\tilde \abf)&=\!\hat a_{\varphi(i_o),0}\!-\!\tilde{a}_{\varphi(i_o),0}\!+\!\hat a_{\varphi(i_o),1}\!-\!\tilde{a}_{\varphi(i_o),1}\nn\\
&\stackrel{(a)}{=}2\hat a_{\varphi(i_o+1),1}-\hat a_{\varphi(i_o),1}-\!\tilde{a}_{\varphi(i_o),1}\nn\\
&\stackrel{(b)}{=}\!\hat a_{\varphi(i_o+1),1}-\hat a_{\varphi(i_o),1}
\end{align}
where $(a)$ is by \eqref{equ:an0_2_i} and $(b)$ is due to \eqref{equ:tilde_a_l}. Similarly, using \eqref{equ:tilde_a_io}\eqref{equ:an0_1}\eqref{equ:an0_2_i} and \eqref{equ:an0_3_i}, we obtain the following for all the other types of $\Dc$
\begin{align*}
\!R_\text{lb}(\Dc;\hat\abf)\!-\!\!R_\text{lb}(\Dc;\tilde \abf)\!&=\!
\begin{cases}
\hat a_{\varphi(i_o+1),1}-\hat a_{\varphi(i_o),1},&\Dc\in\widetilde{\Dc}_{1,1}\\
2(\hat a_{\varphi(i_o+1),1}-\hat a_{\varphi(i_o),1}),&\Dc\in\widetilde{\Dc}_{1,2}\\
\hat a_{\varphi(i_o),1}-\hat a_{\varphi(i_o+1),1}, &\Dc\in\widetilde{\Dc}_{2,1}\\
2(\hat a_{\varphi(i_o),1}-\hat a_{\varphi(i_o+1),1}),&\Dc\in\widetilde{\Dc}_{2,2}\\
0, &\hspace*{-1.5em} \Dc\in\widetilde{\Dc}_{3}\cup\widetilde{\Dc}_4,
\end{cases}
\end{align*}
or more compactly,
\begin{align}\label{equ:diff_1_compact}
&R_\text{lb}(\Dc;\hat\abf)\!-\!\!R_\text{lb}(\Dc;\tilde \abf)=\nn\\
&\hspace{2em}\begin{cases}
j(\hat a_{\varphi(i_o+1),1}-\hat a_{\varphi(i_o),1}),&\Dc\in\widetilde{\Dc}_{1,j}, \; j=1,2\\
j(\hat a_{\varphi(i_o),1}-\hat a_{\varphi(i_o+1),1}),&\Dc\in\widetilde{\Dc}_{2,j}, \; j=1,2\\
0, &  \Dc\in\widetilde{\Dc}_{3}\cup\widetilde{\Dc}_4.
\end{cases}
\end{align}

Substituting \eqref{equ:diff_1_compact} into \eqref{equ:lb_diff}, we have
\begin{align}
&\hspace*{-1em}\bar R_\text{lb}(\hat\abf)-\bar R_\text{lb}(\tilde\abf)\nn\\
=&\sum_{j=1}^{2}\!\sum_{\Dc\in\widetilde{\Dc}_{1,j}}\!\sum_{\dbf\in\Tc(\Dc)}\!\!\!p_{d_1}p_{d_2}j(\hat a_{\varphi(i_o+1),1}-\hat a_{\varphi(i_o),1})\nn\\
&+\sum_{j=1}^{2}\!\sum_{\Dc\in\widetilde{\Dc}_{2,j}}\!\sum_{\dbf\in\Tc(\Dc)}\!\!\!p_{d_1}p_{d_2}j(\hat a_{\varphi(i_o),1}-\hat a_{\varphi(i_o+1),1})\nn\\
=& \; \big(p_{\varphi(i_o)}^2+2\!\!\!\sum_{n'\in \Nc'}\!\!\!p_{n'}p_{\varphi(i_o)}\big)(\hat a_{\varphi(i_o+1),1}-\hat a_{\varphi(i_o),1}) \nn \\
&+\!\big(2\!\!\!\sum_{n'\in \Nc''}\!\!\!p_{n'}p_{\varphi(i_o)}\big) 2(\hat a_{\varphi(i_o+1),1}-\hat a_{\varphi(i_o),1})\nn\\
&+\big(p_{\varphi(i_o+1)}^2\!+2\!\!\!\sum_{n'\in \Nc'}\!\!\!p_{n'}p_{\varphi(i_o+1)}\big)(\hat a_{\varphi(i_o),1}-\hat a_{\varphi(i_o+1),1})\nn \\
&+\!\big(2\!\!\!\sum_{n'\in \Nc''}\!\!\!p_{n'}p_{\varphi(i_o+1)}\big)2(\hat a_{\varphi(i_o),1}-\hat a_{\varphi(i_o+1),1}) \nn\\
\ge& \; 0\label{equ:l_o_finaldiff}
\end{align} where $\Nc' \triangleq \{\varphi(i_o+2),\ldots, \varphi(N)\}$ and $\Nc'' \triangleq \{\varphi(1), \ldots, \varphi(i_o-1)\}$, and the last  inequality is due to the assumption that $p_{\varphi(i_o)}<p_{\varphi(i_o+1)}$ and $\hat a_{\varphi(i_o),1}\ge\hat a_{\varphi(i_o+1),1}$ for $l_o=1$.

\emph{\textbf{Case 2: $l_o=2$.}} From the third case in \eqref{equ:tilde_a_l}, we have
\begin{align}\label{equ_l_o_2_an1}
\hat a_{n,1}=\tilde{a}_{n,1}, \ n\in\Nc.
\end{align}
For $K=2$, $l_o=2$, \eqref{equ:an0_2} and \eqref{equ:an0_3} are respectively given by
\begin{align}
&\hat a_{\varphi(i_o),0}-\tilde{a}_{\varphi(i_o),0}=\hat a_{\varphi(i_o+1),2}-\hat a_{\varphi(i_o),2,}\label{equ:l_o_hati0}\\
&\hat a_{\varphi(i_o+1),0}-\tilde{a}_{\varphi(i_o+1),0}=\hat a_{\varphi(i_o),2}-\hat a_{\varphi(i_o+1),2}.\label{equ:l_o_tildei0}
\end{align}

We  compare $R_\text{lb}(\Dc;\hat \abf)$ and $R_\text{lb}(\Dc;\tilde \abf)$ for different types of $\Dc$'s. For $\Dc\notin\widetilde{\Dc}_{3}$, it is straightforward to show that the expressions of $R_\text{lb}(\Dc;\hat\abf)$ and $R_\text{lb}(\Dc;\tilde\abf)$ are the same as the those for Case 1 ($l_o=1$) in \eqref{equ:lb_a_hat} and \eqref{equ:lb_a_tilde}, respectively. Similar to  Case 1, based on \eqref{equ:lb_a_hat} \eqref{equ:lb_a_tilde} and \eqref{equ_l_o_2_an1} -- \eqref{equ:l_o_tildei0}, except for $\Dc\notin\widetilde{\Dc}_{3}$, we have
\begin{align}\label{equ:R_lb_diff_l2}
&R_\text{lb}(\Dc;\hat\abf)\!-\!\!R_\text{lb}(\Dc;\tilde \abf)\nn\\
&=\begin{cases}
\hat a_{\varphi(i_o+1),2}-\hat a_{\varphi(i_o),2},& \Dc\in\widetilde{\Dc}_{1,j}, \; j=1,\ldots,|\Dc|\\
\hat a_{\varphi(i_o),2}-\hat a_{\varphi(i_o+1),2},& \Dc\in\widetilde{\Dc}_{2,j}, \; j=1,\ldots,|\Dc|\\
0, & \Dc\in\widetilde{\Dc}_4.
\end{cases}
\end{align}
For $\Dc\in\widetilde{\Dc}_3$, we  rewrite \eqref{equ:R_lb_K2} for both $\hat\abf$ and $\tilde\abf$ as follows
\begin{align}
R_\text{lb}(\Dc;\abf)&=a_{\varphi(i_o),0}+a_{\varphi(i_o+1),0}+\max\{a_{\varphi(i_o),1},a_{\varphi(i_o+1),1}\}, \nn\\
& \hspace*{8em} \Dc\in\widetilde{\Dc}_3, \; \abf \in \{\hat{\abf}, \tilde{\abf}\} .\label{equ:l_o_hatD3}
\end{align}
By  \eqref{equ:an0_1}, the sum of the first two terms in \eqref{equ:l_o_hatD3} is the same  for  $\hat\abf$ and $\tilde\abf$. By \eqref{equ_l_o_2_an1}, the third term in \eqref{equ:l_o_hatD3} is identical for  $\hat\abf$ and $\tilde\abf$.
 Thus, we have
\begin{align}\label{equ:l_o2_D3}
R_\text{lb}(\Dc;\hat\abf)-R_\text{lb}(\Dc;\tilde \abf)=0,\quad \Dc\in\widetilde{\Dc}_3.
\end{align}
Following \eqref{equ:l_o_finaldiff}, we substitute \eqref{equ:R_lb_diff_l2} and \eqref{equ:l_o2_D3} into \eqref{equ:lb_diff} and obtain the following \begin{align}
&\bar R_\text{lb}(\hat\abf)-\bar R_\text{lb}(\tilde\abf)\nn\\
&=\Big(p_{\varphi(i_o)}^2+\!\!\!\!\sum_{n'\in \Nc'\cup\Nc''}\!\!\!\!2p_{n'}p_{\varphi(i_o)}\Big) (\hat a_{\varphi(i_o+1),2}-\hat a_{\varphi(i_o),2})\nn\\
&\quad+\Big(p_{\varphi(i_o+1)}^2+\!\!\!\!\sum_{n'\in \Nc'\cup\Nc''}\!\!\!\!2p_{n'}p_{\varphi(i_o+1)}\Big) (\hat a_{\varphi(i_o),2}-\hat a_{\varphi(i_o+1),2})\nn\\
&\ge0\label{equ:l_o_2_finaldiff}
\end{align}
where $\Nc'$ and $\Nc''$ are defined below \eqref{equ:l_o_finaldiff} and the inequality is due to the assumption that $p_{\varphi(i_o)}<p_{\varphi(i_o+1)}$ and $\hat a_{\varphi(i_o),2}\ge\hat a_{\varphi(i_o+1),2}$ for $l_o=2$.

From the above results, we conclude that $\bar R_\text{lb}(\hat\abf)-\bar R_\text{lb}(\tilde\abf)\ge0$, for any $l_o \in \{1,2\}$. This means that, if $p_{\varphi(i_o)}< p_{\varphi(i_o+1)}$, we can always reduce $\bar R_\text{lb}(\hat\abf)$ by switching the values of $\hat a_{\varphi(i_o),l_o}$ and $\hat a_{\varphi(i_o+1),l_o}$. It follows that at the optimality of {\bf P1}, we have $a_{n_1,l_o}\ge a_{n_2,l_o}$,  $l_o=1,2$, for any $n_1,n_2\in\Nc$ satisfying $p_{n_1}\ge p_{n_2}$, \ie the optimal $\abf$ is a popularity-first cache placement. Thus,  {\bf P1} and {\bf P2} are equivalent.
\endIEEEproof

\section{Proof of Lemma \ref{lemma_achi}}\label{Proof:lemma_archi}
\IEEEproof We look at each inner term  $\sum_{\Sc \in \tilde{\Ac}^{l+1}_i}\bar{a}_{\psi(i),l}^\sSc$ of  $R_\text{MCCS}(\dbf;\abf)$  in \eqref{equ:achiev_conv_1}, for $i=1,\ldots,\tilde{N}(\dbf)$. For  cache subgroup $\Ac^{l+1}$, first consider $\tilde{\Ac}^{l+1}_1$, where for any user subset $\Sc\in \tilde{\Ac}^{l+1}_1$,  $\Sc$ includes user $\psi(1)$. Based on the relation of  mappings $\psi(\cdot)$ and $\phi(\cdot)$ discussed above \eqref{a:psi_order}, we have $a_{d_{\psi(1)},l}=a_{\phi(1),l}$, which is the size of the coded message for any user subset $\Sc\in \tilde{\Ac}^{l+1}_1$.  By \eqref{a:psi_order}, \eqref{equ:ach_obj_L3}, and $|\tilde{\Ac}^{l+1}_1|= \binom{K-1}{l}$,
  we have
\begin{align}\label{equ:aver_load_0}
\sum_{\Sc \in \tilde{\Ac}^{l+1}_1}\bar{a}_{\psi(1),l}^\sSc=\binom{K-1}{l}a_{\phi(1),l}.
\end{align}

Denote  $\ddot N_i$ as the number of users that request file $\phi(i)$ but are not in leader group $\Uc$. We have $\ddot N_i\leq N-\tilde{N}(\dbf)$.
For $\tilde{\Ac}^{l+1}_2$ (in which user subsets includes user $\psi(2)$ but not $\psi(1))$,. 
among the total of  $\binom{K-2}{l}$ user subsets, there are $\binom{K-2-\ddot N_1 }{l}$ user subsets that do not contain any user that requests file $\phi(1)$.
The size of coded messages corresponding to these user subsets is ${a}_{d_{\psi(2)},l}=a_{\phi(2),l}$.
For the rest of   $\binom{K-2}{l}-\binom{K-2-\ddot N_1}{l}$ user subsets, since they contain at least one  user $k'$ from the redundant group that requests file $\phi(1)$,  the size of coded message is ${a}_{d_{k'},l}=a_{\phi(1),l}$. Thus, the size of coded message for user subset $\Sc\in \tilde{\Ac}^{l+1}_2$ can be one of the above two cases, and we have 
\begin{align}\label{equ:aver_load_1}
\sum_{\Sc \in \tilde{\Ac}^{l+1}_2}\bar{a}_{\psi(2),l}^\sSc=&\left(\binom{K-2}{l}-\binom{K-2-\ddot N_1}{l}\right)a_{\phi(1),l}\!\nn\\
&+\!\binom{K-2-\ddot N_1}{l}a_{\phi(2),l}.
\end{align}

Following  the similar arguments above, the size of coded message for user subset $\Sc\in \tilde{\Ac}^{l+1}_3$ (\ie including $\psi(3)$ but not $\psi(1)$,$\psi(2)$) can be one of the three types $a_{\phi(1),l}$, $a_{\phi(2),l}$ and $a_{\phi(3),l}$. It follows that
\begin{align}\label{equ:aver_load_2}
&\sum_{\Sc \in \tilde{\Ac}^{l+1}_3}\bar{a}_{\psi(3),l}^\sSc= \left(\binom{K-3}{l}-\binom{K-3-\ddot N_1}{l}\right)a_{\phi(1),l} \nn\\
&\quad\quad+\left( \binom{K-3-\ddot N_1}{l} - \binom{K-3-\ddot N_1-\ddot N_2}{l} \right)a_{\phi(2),l} \nn\\
&\quad\quad+\binom{K-3-\ddot N_1-\ddot N_2}{l} a_{\phi(3),l}.
\end{align}
The first term in the above expression corresponds to the coded messages for the user subsets that contain users from the redundant group  requesting file $\phi(1)$.
The second term is for the coded messages for the user subsets that contain users from the redundant group  requesting file $\phi(2)$ but not $\phi(1)$. The third term represents the coded messages for all the rest user subsets  in $\tilde{\Ac}^{l+1}_2$ that do not request either file $\phi(1)$ or $\phi(2)$.

Following the derivations above,  we can obtain the general expression of  $\sum_{\Sc \in \tilde{\Ac}^{l+1}_i}\bar{a}_{\psi(i),l}^\sSc$  with a recursive pattern. Let $\hat N(i)$ be the total number of redundant requests for  files  $\{\phi(1),\ldots,\phi(i)\}$ (\ie file requests by users in the redundant group). We have $\hat N(i)\triangleq\sum_{j=1}^{i}\ddot N_j$.
Similar to \eqref{equ:aver_load_0}--\eqref{equ:aver_load_2}, for the coded messages for $\Sc \in \tilde{\Ac}^{l+1}_i$, we have
\begin{align}\label{equ:aver_load_3}
&\hspace*{-.3em}\sum_{\Sc \in \tilde{\Ac}^{l+1}_i}\bar{a}_{\psi(i),l}^\sSc=\nn\\
&\left(\binom{K-i}{l}-\binom{K-i-\hat N(1)}{l}\right)a_{\phi(1),l}+\ldots\nn\\
&+\left(\binom{K-i-\hat N(i-2)}{l}-\binom{K\!-i-\hat N(i-1)}{l}\!\right)\!a_{\phi(i-1),l}\!\nn\\
&+\binom{K-i-\hat N(i-1)}{l}a_{\phi(i),l},
\end{align}
for $i=1,\ldots,\tilde{N}(\dbf)$.
Assume that $\hat N(0)=0$. From \eqref{equ:aver_load_0} -- \eqref{equ:aver_load_3}, we have\begin{align}\label{equ:aver_load_5}
\sum_{i=1}^{\tilde{N}(\dbf)}\sum_{\Sc \in \tilde{\Ac}^{l+1}_i}\bar{a}_{\psi(i),l}^\sSc=&\sum_{i=1}^{\tilde{N}(\dbf)}\left[ \sum_{j=i}^{\tilde N(\dbf)}\binom{K-j-\hat N(i-1)}{l}\right.\nn\\
&\left. - \sum_{j=i+1}^{\tilde N(\dbf)}\binom{K-j-\hat N(i)}{l} \right] a_{\phi(i),l}.
\end{align}
Summing up both sides of \eqref{equ:aver_load_5} for $l=0,\ldots,K-1$, we have $R_{\text{MCCS}}(\dbf;\abf)$ as in \eqref{equ:aver_load_4}.
\endIEEEproof

\section{Proof of Theorem \ref{thm:lb_length}}\label{Proof:thm:lb_length}
\IEEEproof
To show that {\bf P4} and {\bf P5} are equivalent for $K=2$,   we will show that $\bar R_\text{lb}(\abf)=\bar R_{\text{MCCS}}(\abf)$, for any given $\abf$. To do so, we only need to compare $R_\text{MCCS}(\dbf;\abf)$ and $R_\text{lb}(\Dc;\abf)$. For $\Kc=\{1,2\}$, we have $|\Dc|=1$ or $2$. We consider the two cases separately below.
\subsubsection{For $|\Dc|=1$} Two users request the same file. We have $d_1=d_2$. Thus, we have $\Dc=\{d_1\}$. By $R_\text{MCCS}(\dbf;\abf)$ in \eqref{CodedMsgPad} and  $R_\text{lb}(\Dc;\abf)$ in \eqref{R_lblb}, it is straightforward to show that
\begin{align}
R_\text{MCCS}(\dbf;\abf)=R_\text{lb}(\Dc;\abf)=a_{d_1,0}+a_{d_1,1}.
\end{align}

\subsubsection{For $|\Dc|=2$} As shown in \eqref{equ:R_lb_K2}, $R_\text{lb}(\Dc;\abf)$ in \eqref{R_lblb} can be written as
\begin{align}
R_\text{lb}(\Dc;\abf)=a_{\xi(1),0}+a_{\xi(2),0}+a_{\xi(1),1}
\end{align}
where  $\xi:\Ic_{|\!\Dc|}\rightarrow\Dc$ is defined as a bijective map such that $a_{\xi(1),1}\ge a_{\xi(2),1}$.

Since two users request different files, we have the leader group $\Uc=\{1,2\}$. For $\Kc=\{1,2\}$,  $R_\text{MCCS}(\dbf;\abf)$ in \eqref{CodedMsgPad} is given by
\begin{align}
R_{\text{MCCS}}(\dbf;\abf) &=\sum_{\Sc\subseteq\{\{1\},\{2\},\{1,2\}\}}\max_{k\in\Sc}a_{d_k,l}\nn\\
&=a_{d_1,0}+a_{d_2,0}+\max\{a_{d_1,1},a_{d_2,1}\}.
\end{align}

By the definition of $\xi:[|\Dc|]\rightarrow\Dc$, we have $R_\text{lb}(\Dc;\abf)=R_{\text{MCCS}}(\dbf;\abf)$.
Thus, we conclude that $\bar R_\text{lb}(\abf)=\bar R_{\text{MCCS}}(\abf)$, and  {\bf P4} and {\bf P5} are equivalent .
\endIEEEproof

\bibliographystyle{IEEEtran}
\bibliography{Yong,IEEEabrv}

\end{document}